\documentclass[1p]{elsarticle}

\journal{J. Logic. Algebr. Program}

\usepackage{tikz}
\usetikzlibrary{positioning}
\usepackage{amsmath}
\usepackage{amsthm}
\usepackage{url}

\usepackage{ifxetex}

\ifxetex
\usepackage{fontspec}
\usepackage{unicode-math}
\else
\usepackage{amssymb}
\usepackage[utf8]{inputenc}
\usepackage[T1]{fontenc}
\newcommand\coloneq{:=}
\fi
\usepackage{booktabs}
\usepackage[noliterate]{maude}
\lstdefinelanguage{membrane}{
	morekeywords=[1]{
		membrane,end,is,ev,cev,xev,pr,signature,ob,obs,var,with,without,import
	},
	alsoletter={.-},
	keywordstyle=[1]{\sffamily\bfseries},
	keywordstyle=[2]{\color{darkgray}\sffamily\bfseries},
	morecomment=[l]{***}
}
\usepackage[colorlinks]{hyperref}
\usepackage[capitalize,noabbrev]{cleveref}

\makeatletter
\let\c@author\relax
\makeatother

\newtheorem{prop}{Proposition}
\newtheorem{lemma}{Lemma}

\newcommand\N{\ensuremath{\mathbb{N}}}

\newcommand\ao{\,\lower1pt\hbox{\normalfont @}\,}

\newcommand*\kywd[1]{\textsf{\bfseries #1}}
\newcommand*\skywd[1]{\textsf{\color{darkgray}\bfseries #1}}

\makeatletter
\def\ps@pprintTitle{\let\@oddhead\@empty
     \let\@evenhead\@empty
     \def\@oddfoot
       {\hbox to \textwidth {\ifnopreprintline\relax\else
        \@myfooterfont \ifx\@elsarticlemyfooteralign\@elsarticlemyfooteraligncenter \hfil\@elsarticlemyfooter\hfil \else \ifx\@elsarticlemyfooteralign\@elsarticlemyfooteralignleft \@elsarticlemyfooter\hfill{}\else \ifx\@elsarticlemyfooteralign\@elsarticlemyfooteralignright {}\hfill\@elsarticlemyfooter \else \normalshape\hfill\begin{tikzpicture}
			\node at (0, 2em) {};
			\node[draw=black!70, fill=black!5, inner sep=5pt, text width=.855\linewidth]{
				Accepted authors' manuscript of the article published in \@journal\ 124 \\
				DOI: \href{https://doi.org/10.1016/j.jlamp.2021.100727}{10.1016/j.jlamp.2021.100727} \hfill License: CC-BY-NC-ND
			};
		\end{tikzpicture} \hfill\fi \fi \fi \fi }
       }\let\@evenfoot\@oddfoot}
\makeatother

\begin{document}

\begin{frontmatter}

\address[ucm]{Facultad de Informática, Universidad Complutense de Madrid, Spain}
\address[itc]{Instituto de Tecnología del Conocimiento, Universidad Complutense de Madrid, Spain}

\cortext[cor1]{Corresponding author}

\title{Simulating and model checking membrane systems using strategies in Maude}
\author[ucm]{Rubén Rubio\corref{cor1}}
\ead{rubenrub@ucm.es}
\author[ucm,itc]{Narciso Martí-Oliet}
\ead{narciso@ucm.es}
\author[ucm]{Isabel Pita}
\ead{ipandreu@ucm.es}
\author[ucm]{Alberto Verdejo}
\ead{jalberto@ucm.es}

\begin{abstract}
	Membrane systems are a biologically-inspired computational model based on the structure of biological cells and the way chemicals interact and traverse their membranes. Although their dynamics are described by rules, encoding membrane systems into rewriting logic is not straightforward due to its complex control mechanisms. Multiple alternatives have been proposed in the literature and implemented in the Maude specification language. The recent release of the Maude strategy language and its associated strategy-aware model checker~\cite{fscd} allow specifying these systems more easily, so that they become executable and verifiable for free. An easily-extensible interactive environment transforms membrane specifications into rewrite theories controlled by appropriate strategies, and allows simulating and verifying membrane computations by means of them.
\end{abstract}


\end{frontmatter}

\section{Introduction}

	A \emph{membrane system} or \emph{P system}~\cite{membraneComputingIntro} is an unconventional distributed and parallel computational model inspired on the structure and interactions of biological cells, proposed in 1998 by Gheorghe Păun. Its theoretical study has led to interesting results like its Turing completeness and the ability to compute NP-complete problems in polynomial time, albeit at an exponential space growth, and its applications cover both biological and non-biological fields~\cite{membraneApplications}. Although simulating P systems is complex, due to its nondeterministic and distributed nature, some simulators have been developed for research and educational purposes~\cite{membraneApplications}. Verification through model checking has also been addressed~\cite{modelCheckingMembranes,exmcPsystem}.

	The connection with rewriting logic and rewriting strategies has been explored in several papers~\cite{exmcPsystem, membraneJournal, andreiCL06, membrane}. These works propose different ways of implementing the membrane control mechanisms in rewriting logic and its specification language Maude~\cite{maude}. In particular, the work in~\cite{membrane} by O. Andrei and D. Lucanu presents a prototype capable of running single evolution steps using a primitive version of the Maude strategy language and some reflective \emph{strategy controllers} that define the control mechanisms dynamically. In this paper, we also specify the membrane control using strategies, but expressed in the stable version of the Maude strategy language, generated at \emph{compile-time} from the membrane specifications, and valid to evaluate them from any possible configuration. The interactive prototype maintains the compatibility with the membrane specification language of~\cite{membrane}, but it is reimplemented and enhanced with new features like loading specifications from file using the new external objects of Maude~3, showing the multiset of rules applied, and computing complete membrane executions. Moreover, we also allow model checking LTL, CTL*, and $\mu$-calculus properties expressed in a builtin but extensible language of atomic propositions. Model checking is directly backed by our model checker for systems controlled by strategies~\cite{fscd,btimemc}, which is able to consider membrane evolution steps as the transitions of the model. Furthermore, the prototype is easily extensible, as illustrated in~\cref{sec:variations} with three common variations of membrane systems, and efficient, as seen in~\cref{sec:performance}. The proposed interactive membrane environment and the strategy-aware model checker can be downloaded from \url{http://maude.ucm.es/strategies}.
	
\subsection{Related work} \label{sec:relatedwork}

Several surveys have been published since the beginning of P systems to compile their huge and growing repertory of simulators~\cite{membraneSoftware,surveyMembranes,superCells2Membranes}. The first prototypes were sequential programs written in Prolog, Lisp, Haskell, or Java that randomly simulate the most basic class of membrane systems, known as transition P systems, the same we address in the main part of this paper. Parallel simulators soon appeared to exploit the intrinsic parallelism of the model, either using multiple threads on the same machine or multiple computers. In addition to standalone programs, libraries have also been proposed to ease the development of variants of membrane systems for specific applications. Indeed, many simulators have been written for concrete biological problems, extending and adapting the formalism to their own features. Probabilistic models, hardware-based implementations using FPGAs, parallel simulations using GPUs~\cite{membraneGPU}, and biochemical realizations have also been explored. In response to the proliferation of simulators for specific purposes, other general-purpose projects like P-Lingua~\cite{plingua} have been proposed. P-Lingua is a programming language and standard for defining P systems, which includes a Java library and has some associated tools like the MeCoSim simulator. Other significant and recent tools are the Infobiotics Workbench~\cite{infobiotics} and kPWorkbench~\cite{kernelPSystem} for kernel P systems, which support model checking of temporal properties through external software like Spin, NuSMV, and PRISM. More details can be found in the surveys cited at the beginning of this paragraph. These references do not mention the prototypes based on rewriting logic~\cite{exmcPsystem, membraneJournal, andreiCL06, membrane} developed by the group of O. Andrei, D. Lucanu, and G. Ciobanu at the Alexandru Ioan Cuza University, in which our work is based. Unlike the prototypes mentioned at the beginning of the section, our goal is showing that strategies are convenient to express the control mechanisms involved in this kind of systems, rather that obtaining an efficient prototype for a specific problem or a general framework where many kinds of P systems can be expressed. However, the flexibility of the rewriting logic framework and the strategy language allows experimenting with different options and configurations easily. 
	
	Outside the field of membrane systems, there are other related biologically-inspired or distributed computational models where the simulation and verification techniques are also subject of active research. For example, this is the case of population protocols~\cite{popop} and chemical reaction networks~\cite{CRNs}.
	
\section{Membrane systems} \label{sec:mems}

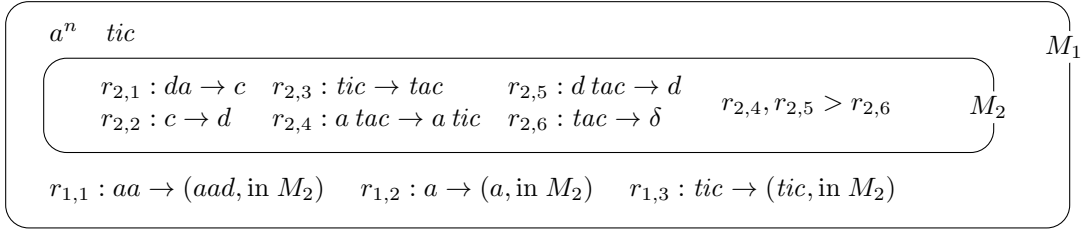
\begin{figure}\centering
\begin{tikzpicture}[rounded corners=2ex, anchor=west, inner sep=2pt]
\draw (0, 1) rectangle (14, 4);
\draw (.5, 2) rectangle (13, 3.25);
\node at (.5, 1.5) {
	$r_{1,1} : aa \to (aad, \mathrm{in}\; M_2)$ \quad
	$r_{1,2} : a \to (a, \mathrm{in}\; M_2)$ \quad
	$r_{1,3} : \mathit{tic} \to (\mathit{tic}, \mathrm{in}\; M_2)$
};
\node at (.5, 3.6) {$a^n \quad \mathit{tic}$};
\node[fill=white] at (13.6, 3.4) {$M_1$};
\node at (1, 2.65) {
	$\begin{array}{lll}
		r_{2,1} : da \to c &
		r_{2,3} : \mathit{tic} \to \mathit{tac} &
		r_{2,5} : d \, \mathit{tac} \to d \\
		r_{2,2} : c \to d &
		r_{2,4} : a \, \mathit{tac} \to a \, \mathit{tic} &
		r_{2,6} : \mathit{tac} \to \delta
	\end{array} \quad r_{2,4}, r_{2,5} > r_{2,6}$
};
\node[fill=white] at (12.6, 2.6) {$M_2$};
\end{tikzpicture}
\caption{Venn diagram of a divisor-calculator membrane system.} \label{fig:venn}
\end{figure}

	Membrane computing~\cite{membraneComputingIntro} is a biologically-inspired computational model where cells are parallel and distributed processing units that communicate by passing objects through their membranes like chemicals traverse that of biological cells. A \emph{membrane system} or \emph{P system} is a collection of cells or membranes populated by a multiset of other nested cells, \emph{objects} playing the role of chemicals, and \emph{evolution rules} describing their reactions and communication. All of them are assumed to be contained inside a single topmost \emph{skin membrane}. Objects are usually opaque identifiers represented by letters, and evolution rules $u \to v$ consist of a multiset $u$ of objects and a multiset $v$ of \emph{targets} of the form $(w, t)$ where $w$ is a multiset of objects and $t$ determines whether these must stay in the membrane ($here$), or be transferred to the enclosing one ($out$) or to a nested one ($in_j$). Moreover, a special symbol $\delta$ causes the enclosing membrane to be dissolved.
Membrane configurations are written like $\langle M_1 \mid a \, b \, c \; \langle M_2 \mid c \, d \rangle \rangle$ where the membrane $M_1$ contains the objects $a$, $b$ and $c$, and the membrane $M_2$, which in turn contains two objects, $c$ and $d$. Formally, the usual definition of a $P$ system with $n$ membranes is a tuple
\[ \Pi = (O, \mu, w_1, \ldots, w_n, R_1, \ldots, R_n, i_o) \]
with the set $O$ of objects, the initial contents $w_i$ and the set of rules $R_i$ of the $n$ membranes, the index $i_o$ of a membrane whose contents or cardinality should be considered as the result of computations, and the initial structure $\mu$ of the nested membranes, usually expressed as a tree or string of paired brackets. Membrane systems are usually represented graphically as Venn diagrams like~\cref{fig:venn}, which describes a system to compute divisors of a number, read as the number of $d$ in the skin membrane. However, both the membrane contents and their structure will likely change during execution, so we will not usually explicit them aside from the configurations themselves. Note that this is a basic definition of P systems, and many variants have been proposed, either including special objects as promoters and inhibitors, allowing membranes to be created or duplicated, using more complex cell topologies like tissue-like and neural-like ones, etc. Some of these variants will be addressed in~\cref{sec:variations}.

	Membrane \emph{computations} are the successive application of \emph{evolution steps}. In turn, evolution steps are the parallel application of as many evolution rules as possible to the objects of each membrane, often regulated by priority relations. Irreducible configurations are those in which no evolution step is possible. More precisely, an evolution step consists of the following phases:
\begin{enumerate}
	\item Applying the evolution rules to each membrane in a \emph{maximal parallel} manner (see below).
	\item Sending and receiving the objects contained in $out$, $in$, and $here$ targets.\item Dissolving membranes containing $\delta$, thus dropping its objects and membranes to the enclosing membrane.
\end{enumerate}
The \emph{maximal parallel rewrite} step is described by a multiple choice of rules or multiset $A_i : R_i \to \N$ for each membrane $M_i$. A multiset of rules $A$ \emph{can be applied} to a multiset of objects $W$ if the union of their left-hand sides with multiplicities is contained in $W$, and the result has that union replaced by the union of the right-hand sides in $A$. Such a choice $A$ is \emph{maximal} if $A + \{ r \}$ cannot be applied to $W$ for no matter which rule $r$.\footnote{For multisets, we write $(A + B)(r) = A(r) + B(r)$, $(A - B)(r) = \max \{ A(r) - B(r), 0 \}$, $A[r / n](r) = n$ and $A[r /n](r') = A(r')$ if $r \neq r'$, and $\{r\}$ for the multiset with a single $r$.} In summary, a maximal parallel rewrite is the application of a maximal multiset of rules to each membrane $(A_1, \ldots, A_n)$ with at least one non-empty $A_i$. The choice of each $A_i$ may not be unique, so this phase is nondeterministic.

	Moreover, when a priority relation $\rho_i$ is imposed, not all choices are \emph{admissible}. Two ways of understanding rule priorities are considered:
\begin{itemize}
	\item a \emph{weak} sense, in which a choice $A_i$ is admissible if for all $r \in R_i$ either $A_i(r') = 0$ for all $r >_{\rho_i} r'$ or the choice $A_i[r'/ 0]_{r >_{\rho_i} r'} + \{r\}$ cannot be applied.
	\item and a \emph{strong} sense, in which a choice $A_i$ is admissible if it is admissible in the weak sense and, in addition, $A_i(r') = 0$ for all $r >_{\rho_i} r'$ such that $A_i(r) > 0$;
\end{itemize}
Intuitively, the membrane objects present in the configuration at each step should be distributed among the membrane rules according to their priority relation. A rule should only be assigned objects if they cannot be used for higher priority rules. Using the objects left by them is possible in the weak sense, but disallowed in the strong sense. The parallel application of evolution rules is well-defined as a multiset, because the order in which they are applied does not matter, since they only subtract chemicals from the membrane multiset. A relevant consequence is that the maximal parallel application of rule priorities in the weak sense can be calculated by the exhaustive sequential application of the rules where priorities are considered locally at each step.

	For example, the divisor calculator of~\cref{fig:venn} is intended to be executed from the initial configuration $\langle M_1 \mid a^n \, \mathit{tic} \; \langle M_2 \mid \; \rangle \rangle$ where $a^n$ means $n$ copies of $a$. The maximal parallel application of the rules in $M_1$ transfers in a single evolution step all the $a$ and the $\mathit{tic}$ objects to $M_2$ along with a nondeterministic number of $d$ between $0$ and $n/2$, which is the divisor candidate. In fact, objects are communicated in the second phase of the evolution step according to the $\mathrm{in}\; M_2$ targets. The next steps take place in $M_2$, where the number of $a$s is divided by the number of $d$s using successive subtractions that take two evolution steps each. When the $\mathit{tic}$ object is present, $r_{2,1}$ removes an $a$ for each $d$ yielding a $c$, and $r_{2,3}$ turns the $\mathit{tic}$ into a $\mathit{tac}$. In the next evolution step, the missing $d$s are recovered with $r_{2,2}$ and the rest of the subtraction is checked. If there are $a$ objects left from the previous step, the division is not completed yet and $r_{2,4}$ transforms $\mathit{tac}$ to $\mathit{tic}$ for a new iteration. If there are $d$ objects left, the number of $a$s was not a divisor of the number of $d$s, so the execution is stopped by removing the $\mathit{tac}$ object with $r_{2,5}$. Otherwise, there is neither $a$s nor $d$s in the configuration, so the division is exact and a true divisor has been found. Then, the rule $r_{2,6}$ introduces the symbol $\delta$ to trigger the dissolution of $M_2$ in the third phase of the evolution step, whose objects are dropped to $M_1$. Note that $r_{2,6}$ can only be applied according to the priorities if $r_{2,4}$ and $r_{2,5}$ cannot be applied, so that the membrane is not dissolved unless a divisor has been found. In this case, weak and strong priorities would produce the same result, since the application of $r_{2,4}$ or $r_{2,5}$ consumes the $\mathit{tac}$ symbol, which impedes the execution of $r_{2,6}$.

\section{Rewriting logic, Maude and its strategy language}

	\emph{Rewriting logic}~\cite{rewritingLogic} was proposed by J. Meseguer as a unified model of concurrency where nondeterministic and possibly conditional rewrite rules are defined on top of an equational logic. A rewrite theory $\mathcal R = (\Sigma, E, R)$ consists of a signature $\Sigma$ of sorts and operators, a set of equations $E$, and a set of rewrite rules $R$. Signature and equations typically describe the static part of a model, while rules represent the nondeterministic concurrent changes it may suffer.
The executions of a rewriting system are the successive and independent application of these rules on terms, modulo equations and axioms like commutativity, associativity, and identity. 

Maude~\cite{maude} is a specification language based on rewriting logic, where rewrite systems can be specified compositionally, executed, and analyzed. Specifications are written in a mathematical-like notation and organized in modules of different kinds: functional modules (\kywd{fmod}) represent equational theories with declarations of \kywd{sort}s, \kywd{subsort} relations, and \kywd{op}erators. Beside their signature, operator declarations may include some attributes between brackets that specify the structural axioms and other features applied to them. Moreover, functional modules may include (possibly conditional) equations of the form:
\[ [\kywd{c}]\kywd{eq} \quad l \;\texttt{=}\; r \quad [\; \kywd{if}\; \bigwedge_i l_i \;\texttt=\; r_i \;\;\verb|/\|\;\; \bigwedge_i l'_i \;\texttt{:=}\; r'_i \;\;\verb|/\|\;\; \bigwedge_i t_i \;\texttt{:}\; s_i \;]\; \texttt. \]
Equations are applied as if they were oriented from left to right on any position where they match.
Conditional equations are introduced by the \kywd{ceq} instead of \kywd{eq} keyword, and its condition fragments are satisfied when their term pairs coincide modulo equations and axioms (potentially instantiating left-hand side variables by matching in the \texttt{:=} variant), and when the terms $t_i$ belong to the sorts $s_i$. For example, the following functional module specifies multisets of integer numbers:
\begin{lstlisting}
fmod MULTISET is
  protecting INT .
  sort Multiset .
  subsort Int < Multiset .

  op empty : -> Multiset [ctor] .
  op __    : Multiset Multiset -> Multiset 
               [ctor assoc comm id: empty] .

  var N : Int . vars U V : Multiset .

  op contains : Multiset Multiset -> Bool .
  eq contains(empty, V) = true .
  eq contains(N U, N V) = contains(U, V) .
  eq contains(U, V) = false [owise] .
endfm
\end{lstlisting}
Underscores in operator names mark the gaps where arguments are entered, except in those written in prefix notation. In the module above, \texttt{Int} is made a subsort of \texttt{Multiset}, the juxtaposition operator \texttt{\_\_} is declared associative, commutative, and having \texttt{empty} as identity, and the module \texttt{INT} is imported. The Maude prelude provides some modules like \texttt{INT} that specify integer and floating-point numbers, strings, lists, sets, etc. Importation can be done with the keywords \texttt{protecting}, \texttt{extending}, or \texttt{including} that declare whether the definitions of the imported module will be kept unchanged, extended, or modified arbitrarily. The \texttt{contains} predicate is defined using three equations to decide whether its first argument is contained in the second. Equations marked with \texttt{owise} are executed only after all equations without this attribute have failed.

	System modules (\kywd{mod}) are complete rewrite theories and declare rewrite rules
\[ [\kywd{c}]\kywd{rl} \quad l \;\texttt{=>}\; r \quad [\; \kywd{if}\; C \;\verb|/\|\; \bigwedge_i l_i \;\texttt{=>}\; r_i \;]\; \texttt. \]
Conditional rules may include rewriting conditions in addition to those available for equations, where terms matching $r_i$ are searched by rewriting with rules from $l_i$. Every possible match of the left-hand side and the condition yields a different application of the rule. Continuing with the example, the following module \texttt{MULTISET-RLS} extends the previous \texttt{MULTISET} with two rules, \texttt{sum} and \texttt{add}, that respectively take two numbers and replace them by their sum, or increment a number by a given fixed amount.
\begin{lstlisting}
mod MULTISET-RLS is
  protecting MULTISET .

  vars N M K : Int .

  rl [sum] : N M => N + M .
  rl [add] : N => N + K [nonexec] .
endm
\end{lstlisting}
Note that the \texttt{add} rule contains an unbounded variable \texttt{K} in its right-hand side. What would be an error without the \texttt{nonexec} attribute, which excludes the rule from being applied, can be useful when combined with a strategy language able to instantiate this variable. Using this strategy language, discussed in the next section, strategy modules (\kywd{smod}) specify alternative ways of applying these rules.

	The Maude system offers several commands to \texttt{reduce} terms equationally, to \texttt{rewrite} a term with the rules modulo equations and axioms, to \texttt{search} terms matching patterns in the rewriting graph, etc. More information can be found in the Maude manual~\cite{maude}.

\subsection{The Maude strategy language}

	Rewriting strategies have been used in Maude since its beginnings thanks to its reflective features, explained in~\cref{sec:reflection}. However, writing and understanding (usually verbose) metalevel programs to control rewriting is a complex task, and so an object-level strategy language inspired on this experience and on previous strategy languages like ELAN~\cite{elan} and Stratego~\cite{stratego} has been proposed. After being prototyped in Full Maude and tested, the language was finally implemented at the C++ level in Maude~3 with new features like compositional and parameterized strategy modules~\cite{pssm}. Expressions of the strategy language restrict the possible evolution of a rewriting system, and they can be formally seen as transformations from an initial state to the set of terms yielded by this controlled but not necessarily deterministic rewriting. Two Maude commands \texttt{srewrite} and \texttt{dsrewrite} explore all possible execution paths (the second using a depth-first search) to show this set of solutions.

	The main ingredient of the language is the application of rules \texttt{$rl$[$x_1$ \textleftarrow{} $t_1$, $\ldots$ $x_n$ \textleftarrow{} $t_n$]\{$\alpha_1$, $\ldots$, $\alpha_m$\}} referred by their labels $rl$ and taking an optional initial substitution, which is applied to both sides of the rule and its condition before matching. If the rule to be applied includes rewriting conditions, a comma-separated list of strategies must be given between curly brackets to control all of them. Rules are applied anywhere within the term by default, but its application can be restricted to the top with the \skywd{top}\texttt{($\alpha$)} combinator.
Tests \skywd{match $P$ s.t.\ $C$} discard executions unless the subject term matches $P$ and satisfies the condition $C$. The initial keyword can be changed to \skywd{amatch} to match anywhere within the term. These elements can be combined with the concatenation $\alpha \texttt; \beta$ that executes $\beta$ on the results of $\alpha$, the disjunction $\alpha \texttt| \beta$ that permits the executions allowed by any of its arguments, the iteration $\alpha \texttt*$ that iterates $\alpha$ any number of times, and the conditional $\alpha \texttt? \beta \texttt: \gamma$ that evaluates $\alpha$ and then $\beta$ on its results, but if $\alpha$ does not produce any, it executes $\gamma$ on the initial term. Two constants \skywd{idle} and \skywd{fail} represent the strategy that produces the initial term as result and the strategy that does not produce any result at all. In general, we say that any strategy $\alpha$ \emph{fails} in this latter case.  Another combinator allows rewriting selected subterms \skywd{matchrew $P$ s.t. $x_1$ using $\alpha_1$, $\ldots$, $x_n$ using $\alpha_n$}. The terms matched by the variables $x_1, \ldots, x_n$ in the pattern are rewritten in parallel using $\alpha_1, \ldots, \alpha_n$ respectively, and their results are combined to produce the global results. In addition to these basic combinators, some other derived ones are included in the language. The \lstinline[mathescape, basicstyle=\ttfamily]{$\alpha$ or-else $\beta$} combinator will play an important role for dealing with priorities in this paper, since it evaluates to the results of $\alpha$ except in case they are none, where it takes the results of $\beta$. It is defined as \lstinline[mathescape, basicstyle=\ttfamily]{$\alpha$ ? idle : $\beta$}. The normalization operator $\alpha \texttt!$ is defined by \lstinline[basicstyle=\ttfamily, mathescape]{$\alpha$ * ; ($\alpha$ ? fail : idle)}, so it executes $\alpha$ as many times as possible.

	Moreover, strategies can be given names to be called, receive parameters, and be defined recursively in strategy modules. These modules, declared with the \kywd{smod} keyword, may import modules of any kind and include strategy declarations \lstinline[mathescape, basicstyle=\ttfamily]|strat $\mathit{name}$ : $s_1 \;\cdots\; s_n$ @ $s$ .| with the signature of its parameters and the sort $s$ of the intended terms where it will be applied, and definitions \lstinline[mathescape, basicstyle=\ttfamily]|sd $\mathit{name}$($p_1$, $\ldots$, $p_n$) := $\alpha$ .| where $p_1, \ldots, p_n$ are patterns containing variables that can be used in the strategy expression. Conditional definitions are introduced with the \kywd{csd} keyword and support the same conditions as equations. Strategies are called with \texttt{$\mathit{name}$($t_1$, $\cdots$, $t_n$)}, and all definitions whose left-hand side matches the call will be executed. For example, the following strategy module declares and defines a strategy \texttt{increment} that increments by the given number all the elements of the multiset.
\begin{lstlisting}
smod MULTISET-STRATS is
  protecting MULTISET-RLS .

  strat increment : Nat @ Multiset .

  vars N K : Int . var U : Multiset .

  sd increment(K) := match empty | matchrew N U
                     by N using top(add[K <- K]),
                        U using increment(K) .
endsm
\end{lstlisting}
Its definition is a disjunction of two exclusive cases: the subject term may be either an \texttt{empty} multiset matched by the test, or a non-empty one decomposed in an element and the rest by the recursive \texttt{matchrew}.
The \texttt{srewrite} command can then be used to rewrite using this strategy:
\begin{lstlisting}[language={}]
Maude> srewrite 1 2 3 4 5 using increment(1) .

Solution 1
rewrites: 650
result Multiset: 2 3 4 5 6

No more solutions.
\end{lstlisting}
More details about the strategy language can be found in~\cite[\S 10]{maude}.

\subsection{Reflection and metalanguage interfaces} \label{sec:reflection}

	Rewriting logic is a reflective logic, whose objects and operations can be consistently represented in itself. Maude offers a predefined \emph{universal theory}~\cite[\S 17]{maude} to metatheoretically represent terms, equations, rules, modules, and so on. Operations like matching, reduction, and rule application can be programmed generically in this theory using equations, but Maude provides special operators backed by the object-level implementation in C++ to allow efficient reflective computations. Metarepresentations can in turn be metarepresented and terms be moved between different levels, thus yielding arbitrarily high reflective towers.

	This universal theory is specified in \texttt{META-LEVEL} and its imported modules, and it relies on the \texttt{Qid} sort of \emph{quoted identifiers}, arbitrary words prefixed by an apostrophe, to represent indivisible elements like names, sorts, variables, and constants. Composite elements are constructed using Maude operators, like terms are with the operator \texttt{\_[\_] : Qid NeTermList -> Term}, so that \verb|'_+_['X:Nat, 's_['0.Zero]]| is the representation of \texttt{X + s~0}. The metarepresentation of strategy expressions faithfully reproduces its object-level syntax in most cases, and they are specified as terms of the \texttt{Strategy} sort. For instance, a simple rule application is written \texttt{'label[none]\{empty\}}, and a strategy call \texttt{'name[[$\mathit{TL}$]]} with $\mathit{TL}$ a possibly \texttt{empty} list of metarepresented terms. Operator and strategy declarations, equations, rules, strategy definitions, and similar statements are also represented as terms with a syntax similar to the object-level reference. They usually appear in the metarepresentation of modules, terms with argument slots like \verb|smod_is_sorts_._______endsm| for each kind of module component, which can be obtained by some auxiliary functions \texttt{getOps}, \texttt{getEqs}, \texttt{getRls}, etc.

	Operations are accessible through some \emph{descent functions} like \texttt{metaMatch} for matching, \texttt{metaApply} for rule application, \texttt{metaReduce} for equational reduction, \texttt{metaRewrite} for rule rewriting, etc. The \texttt{srewrite} and \texttt{dsrewrite} commands are accessible through the \texttt{metaSrewrite} descent function that allows enumerating the results of rewriting a term using a strategy in certain module, all of them given at the metalevel. Another useful descent function for building metalanguage interfaces is \texttt{metaParse} that parses terms on a given module and sort.
\begin{lstlisting}
op metaParse : Module VariableSet QidList Type? 
               ~> ResultPair? [special (...)] .
\end{lstlisting}
On success, it returns a pair with the metarepresentation of the term and its least sort. Its input should be a list of tokens of sort \texttt{QidList}, which can be obtained from a string using the \texttt{tokenize} function. Ad-hoc grammars can be expressed as Maude modules to parse arbitrary languages, with the possibility of including unparsed fragments known as \emph{bubbles} within its terms for more complex multilayered parsing.
The complete specification of the metalevel is included in the Maude prelude and explained in~\cite[\S 17]{maude}.

	Moreover, the interactive capabilities of Maude have been enhanced in its 3 version due to new \emph{external objects} that allow reading and writing files as well as the standard input and output streams. External objects are an object-oriented mechanism that allow Maude programs to communicate with the outside word, already used in previous versions for Internet sockets. The standard \texttt{CONFIGURATION} module defines an extensible signature for defining objects and messages, which are held in a common soup or multiset where objects read and introduce messages by means of rewrite rules. The command \texttt{erewrite} conducts rewriting of these configurations following an object-fair strategy and handling the messages issued to and by the implicit external objects. In this case, the \texttt{STD-STREAM} and \texttt{FILE} modules in the \texttt{file.maude} file of the Maude distribution declare the \texttt{stdin}, \texttt{stdout}, \texttt{fileManager}, and \texttt{file($n$)} objects, and the messages \texttt{getLine}, \texttt{read}, \texttt{write}, \texttt{openFile}, \texttt{close}, etc., that are sent to and received from them.\footnote{In previous versions, Maude offered a different input/output facility called \texttt{LOOP-MAUDE}, which is currently deprecated in favor of this more flexible method.} The main representative of an interactive interface leveraging the reflective features of Maude is Full Maude~\cite[Part II]{maude}, an extensible Maude interpreter where many currently stable features have been first tested.

\section{Representing membrane systems in Maude} \label{sec:repr}

	Membrane systems have already been specified in rewriting logic and in Maude~\cite{membraneJournal,membrane,exmcPsystem,memrewlogPromoters}. Multisets of objects and the nested membrane structure are naturally represented by terms with commutative and associative operators, but the challenge is applying evolution rules locally and in a maximal parallel way. This has been solved in previous works by
\begin{itemize}
	\item representing evolution rules as rewrite rules and controlling their application with the Maude reflective features~\cite{exmcPsystem},
	\item by representing evolution rules as data and computing the steps at the object level~\cite{membraneJournal},
	\item or even using a particular combination of reflection and a primitive version of the Maude strategy language~\cite{membrane}. \end{itemize}
In either case, the specification of membrane configurations is very similar, and ours only slightly differs from those:
\begin{lstlisting}
mod P-SYSTEM-CONFIGURATION is
  including QID-LIST .

  sorts Obj Membrane MembraneName Target TargetMsg .
  sorts EmptySoup MembraneSoup ObjSoup TargetSoup Soup .

  subsort  Obj < ObjSoup .
  subsort  Membrane < MembraneSoup .
  subsort  TargetMsg < TargetSoup .
  subsorts EmptySoup < MembraneSoup ObjSoup TargetSoup < Soup .

  op <_|_>   : MembraneName Soup -> Membrane [ctor] .
  op delta   : -> Obj [ctor] .

  op empty   : -> EmptySoup [ctor] .
  op __      : Soup Soup -> Soup [ctor assoc comm id: empty] .
\end{lstlisting}
A membrane is identified by a name and contains a multiset of juxtaposed objects, membranes, and targets of sort \texttt{Soup}. Each component defines a subsort, \texttt{ObjSoup}, \texttt{MembraneSoup}, \texttt{TargetSoup}, to facilitate operating with them. Several omitted operator declarations specify how these subsorts combine with the \texttt{\_\_} operator. Target messages are expressed as pairs:
\begin{lstlisting}
  ops here out : -> Target [ctor] .
  op  in_ : MembraneName -> Target [ctor] .
  op `(_,_`) : Soup Target -> TargetMsg [ctor frozen (1)] .
\end{lstlisting}
Rewriting is proscribed in the first argument of the message with the \texttt{frozen (1)} annotation, since objects in messages have been generated by evolution rules and must not be used until the next evolution step.
Messages with the same target are combined into a common pair by an equation, and three rules are defined to resolve the communication between cells:
\begin{lstlisting}
  vars MN MN'  : MembraneName .
  vars W W' CW : Soup .
  var  T       : Target .

  eq (W, T) (W', T) = (W W', T) .

  rl [here] : (W, here) => W .
  rl [in]   : (CW, in MN) < MN | W > => < MN | W CW > .
  rl [out]  : < MN | W (CW, out) > => < MN | W > CW .
\end{lstlisting}
Finally, another rule \texttt{dis} triggers the effect of the $\delta$ symbol by dissolving the non-skin membrane where it is contained. The skin or outermost membrane is never dissolved, as enforced by the nested pattern in the rule.
\begin{lstlisting}
  rl [dis] : < MN | W < MN' | W' delta > > => < MN | W W' > .
endm
\end{lstlisting}
This \texttt{P-SYSTEM-CONFIGURATION} module is the common and generic base of the rewriting theories that will specify concrete membrane systems, where the \texttt{MembraneName} and \texttt{Obj} sorts are populated, and the evolution aspects are defined.

	Evolution rules are represented as identical rewrite rules, delegating their controlled application to strategies. These strategies are partially generic and partially dependent on the rules and priorities of the membrane system, but not on any particular configuration. For a membrane $M$ with rules $r_1, \ldots, r_n$ and without priorities, its specific strategy definition will be\footnote{Another possibility to limit the locality of evolution rules is adding a membrane context: the rule $v \to w$ for the membrane $M$ could also be transformed into $\texttt{<} \;M\; \texttt{|} \;v\;\, \texttt{S:Soup >} \Rightarrow \texttt{<} \;M\; \texttt{|} \;w\;\, \texttt{S:Soup >}$.}
\begin{lstlisting}[mathescape]
sd membraneRules($M$) := $r_1$ | $\cdots$ | $r_n$ .
\end{lstlisting}
The generic part is specified in the following module \texttt{P-SYSTEM-STRATEGY}, where the strategy \texttt{mpr} defines the maximal parallel step, in terms of the system-specific \texttt{handleMembrane}. \begin{lstlisting}
smod P-SYSTEM-STRATEGY is
  protecting P-SYSTEM-CONFIGURATION .

  strat  handleMembrane : MembraneName @ Soup .
  strats mpr visit-mpr communication @ Soup .
  strat  nested-mpr : MembraneSoup @ Soup .

  var MN : MembraneName .  var TM : TargetMsg .
  var S  : ObjSoup .       var TS : TargetSoup .
  var MS : MembraneSoup .  var K  : Nat .

  sd mpr := visit-mpr ; amatch TM ;
            communication ; (dis !) .

  sd communication := (in | out | here) ! .

  sd visit-mpr := matchrew < MN | S MS > 
                   by S  using handleMembrane(MN),
                      MS using nested-mpr(MS) .

  sd nested-mpr(empty) := idle .
  sd nested-mpr(M MS) :=  (matchrew M MS 
       by M using visit-mpr) ; nested-mpr(MS) .
\end{lstlisting}
The three phases of an evolution step (see~\cref{sec:mems}) are concatenated in the main strategy \texttt{mpr}: (1) \texttt{visit-mpr} applies the evolution rules to the membranes, (2) \texttt{communication} transmits all targeted objects through the membranes, and (3) \texttt{dis !} dissolves them exhaustively. The \texttt{visit-mpr} strategy executes the parameter \texttt{handleMembrane} on the objects of the topmost membrane, and then continues with the nested ones using \texttt{nested-mpr}. The requirement that at least an evolution rule must be applied in the whole system is enforced by the \lstinline[basicstyle=\ttfamily]{amatch TM} test that discards an execution if no target has been generated in the whole configuration.\footnote{In previous versions, \texttt{visit-mpr} and their auxiliary strategies take care themselves of whether at least a rule has been applied by using conditional operators and observing whether rules succeeded. However, looking at the presence of a target message is equivalent and simpler.}
\texttt{nested-mpr} receives the full nested membrane soup and recursively processes the membranes one at a time. Note that the \texttt{nested-mpr} strategy is not efficient, since all possible matches of the set-like argument will be tried at each call, hence unnecessarily processing the membranes in all possible orderings. This will be prohibitive when dealing with exponential-size membranes in~\cref{sec:division}, where alternatives are discussed.

	The \texttt{handleMembrane} strategy can be accommodated to different situations, depending on the rule priorities and their interpretation. The case without priorities or with their weak sense can be handled by the following strategy \texttt{inner-mpr}:
\begin{lstlisting}
  strat inner-mpr : MembraneName @ Soup .
  sd inner-mpr(MN) := membraneRules(MN) ! .
\end{lstlisting}
For a membrane name \texttt{MN}, it calls \texttt{membraneRules} repeatedly until it cannot be applied again. Since the multiset argument of target messages is declared \texttt{frozen}, i.e.\ it is excluded from rewriting, products of the evolution rules are not used in the same step.\footnote{An alternative to exclude rule products from being used again in the same step is separating loose objects from targets with a \texttt{matchrew} pattern \texttt{OS TS}. However, this is slower.}
With \texttt{membraneRules} being a disjunction of rules, as above, the \texttt{mpr} strategy implements a maximal parallel step. 

\begin{prop}
	The strategy \texttt{mpr} executes a maximal parallel evolution step without priorities when \texttt{membraneRules($M$)} is defined as the disjuntion of all rules for $M$, i.e., under these conditions, a configuration $C'$ is obtained by an evolution step from $C$ iff $C'$ is a result of the strategy \texttt{mpr} applied on $C$.
\end{prop}

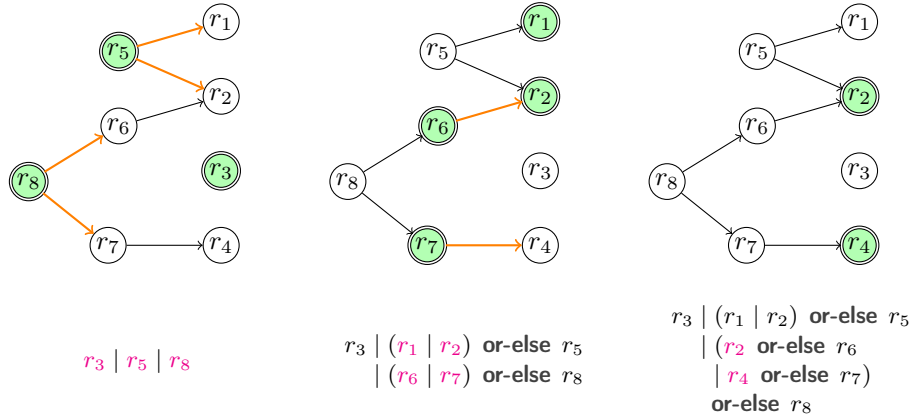
\begin{figure}[h]\centering
\newcommand*\lattice[7]{
	\node[rule, #3] at (#5)                            (R1#1) {$r_1$};
	\node[rule, below=of R1#1, #3, #4]                 (R2#1) {$r_2$};
	\node[rule, below=of R2#1, #2]                     (R3#1) {$r_3$};
	\node[rule, below=of R3#1, #4]                     (R4#1) {$r_4$};
	\node[rule, above left=0.25cm and 1cm of R2#1, #2] (R5#1) {$r_5$};
	\node[rule, above left=0.25cm and 1cm of R3#1, #3] (R6#1) {$r_6$};
	\node[rule, left=of R4#1, #3]                      (R7#1) {$r_7$};
	\node[rule, above left=.5cm and .7cm of R7#1, #2]  (R8#1) {$r_8$};

	\draw[->, #6] (R5#1) -- (R1#1);
	\draw[->, #6] (R5#1) -- (R2#1);
	\draw[->, #7] (R6#1) -- (R2#1);
	\draw[->, #7] (R7#1) -- (R4#1);
	\draw[->, #6] (R8#1) -- (R6#1);
	\draw[->, #6] (R8#1) -- (R7#1);
}

\newcommand*\actrule[1]{{\color{magenta}r_#1}}
\newcommand\orelse{\,\;\skywd{or-else}\;\,}

\begin{tikzpicture}[rule/.style={circle,draw, inner sep=1pt}, node distance=.5cm and 1cm]
\lattice{0}{double, fill=green!30}{}{}{0, 0}{thick, draw=orange}{}
 	 \lattice{1}{}{double, fill=green!30}{}{4.2cm, 0}{}{thick, draw=orange}
  	 \lattice{2}{}{}{double, fill=green!30}{8.4cm, 0}{}{}

	\node at (-1.1, -4.5) {\small $\actrule3 \mid \actrule5 \mid \actrule8$};
	\node at (3.2, -4.5) {\small $\begin{array}{r@{\;}l}
		r_3 &\mid (\actrule1 \mid \actrule2) \orelse r_5 \\
		&\mid (\actrule6 \mid \actrule7) \orelse r_8
	\end{array}$};
	\node at (7.5, -4.5) {\small $\begin{array}{r@{\;}l}
		r_3 &\mid (r_1 \mid r_2) \orelse r_5 \\
		&\mid (\actrule2 \orelse r_6 \\
		&\kern5pt\mid \actrule4 \orelse r_7) \\
		&\orelse r_8
	\end{array}$};
\end{tikzpicture}
\caption{Strategy generation for an example weak priority relation.} \label{fig:priogen}
\end{figure}

	Priorities in the weak sense can also be handled by adding a lattice of \skywd{or-else} combinators to the \texttt{membraneRules} definition. Assuming that $P \subseteq R \times R$ is a generator set for this priority relation, the following recursive procedure is able to generate a strategy respecting the priorities at each application:
\begin{enumerate}
	\item consider the disjunction \lstinline[mathescape, basicstyle=\ttfamily]@$r_1$ | $\cdots$ | $r_n$@ of the minimal elements in $P$,
	\item replace each such $r$ by \lstinline[mathescape, basicstyle=\ttfamily]@($r'_1$ | $\cdots$ | $r'_n$) or-else $r$@ where $r'_1, \ldots, r'_n$ are all rules satisfying $(r'_i, r) \in P$, and
	\item iterate (2) on the newly introduced rules up to the maximal elements.
\end{enumerate}
The algorithm can be optimized by simplifying \lstinline[mathescape, basicstyle=\ttfamily]@$\alpha$ or-else $\beta$ | $\alpha$ or-else $\gamma$@  on the fly to \lstinline[mathescape, basicstyle=\ttfamily]@$\alpha$ or-else ($\beta$ | $\gamma$)@. The correctness of this procedure follows from the fact that a rule will only be applied when its predecessors in the order have failed, due to the semantics of the \skywd{or-else} combinator, and that all rules appear in the expression because they must be reachable from the minimal elements of the order relation. The size of the strategy is bounded by the number of rules and the relation pairs in $P$. An example is shown in~\cref{fig:priogen}.

\begin{prop}
	The strategy \texttt{mpr} executes a maximal parallel evolution step with weak priorities when \texttt{membraneRules($M$)} is defined as indicated above.
\end{prop}

	When the priority of the rules is understood in the strong sense, the skeleton of \texttt{inner-mpr} cannot be exploited since it executes each rule independently in the sequence, and respecting strong priorities requires knowing which rules have already been applied. However, a variation of the previous procedure can be used. In this case, the parameter \texttt{handleMembrane} is defined using a new recursive strategy \texttt{strong-mpr} that receives a set of labels of the rules that have already been applied:
\begin{lstlisting}
  strat strong-mpr : MembraneName QidSet @ Soup .
  sd handleMembrane(MN) := strong-mpr(MN, empty) .
\end{lstlisting}
The definition of \texttt{strong-mpr($M$, AR)} for a membrane $M$ whose priority is generated by $P$ is given recursively as:
\begin{enumerate}
	\item Take the disjunction of $\alpha_r \coloneq$ \lstinline[mathescape]@$r$ ; strong-mpr($M$, ($r$, AP))@ for every minimal element of~$P$. The recursive call adds $r$ to the comma-separated set \texttt{AP} of applied rules.
	\item Replace each element $\alpha_r$ in the disjunction by
\begin{lstlisting}[mathescape]
($\alpha_{r_1}$ | $\cdots$ | $\alpha_{r_n}$) or-else
 (match S s.t. $\{ r' \in R_i \mid r' >_{\rho_i} r \}$ intersect AP = empty ; $\alpha_r$)
\end{lstlisting}
where \texttt{S} is a variable of sort \texttt{Soup}, and $r_1$, \ldots, $r_n$ are all the elements that satisfy $(r_i, r) \in P$. The test prevents the rule $r$ from being applied if a rule with higher priority has already been used.
	\item Iterate (2) on the newly introduced elements up to the maximal ones.
	\item Take $\alpha \;\skywd{or-else}\; \skywd{idle}$ where $\alpha$ is the result of (3).
\end{enumerate}
Because of the guards, the previous construction guarantees that rules are executed only if no rule with higher precedence has been applied.

\begin{prop}
	The strategy \texttt{mpr} executes a maximal parallel evolution step with strong priorities when \texttt{handleMembrane} is instantiated to the \texttt{strong-mpr} strategy described above.
\end{prop}

	Finally, executions up to irreducible configurations are described by the following strategy \texttt{mcomp}. Unlike \texttt{mcomp2}, used to define it, trivial executions that do not take any step are excluded.
\begin{lstlisting}
  strats mcomp mcomp2 @ Soup .
  sd mcomp  := mpr ; mcomp2 .
  sd mcomp2 := mpr ? mcomp2 : idle .
\end{lstlisting}
Computations up to a maximum number of steps or until a given number of objects is reached can also be specified with definitions like
\begin{lstlisting}
  sd mcomp(0) := idle .
  sd mcomp(s(K)) := mpr ? mcomp(K) : idle .
  sd mcomp-obj(K) := match S s.t. numObjsRec(S) >= K
                       or-else (mpr ? mcomp-obj(K) : idle) .                
endsm
\end{lstlisting}
where \texttt{numObjsRec} recursively counts the number of objects in the given soup.

	Fortunately, the interactive environment in~\cref{sec:environ} will make the manual instantiation of these strategies unnecessary: \texttt{membraneRules} and \texttt{strong-mpr} are constructed equationally at the metalevel following the above procedures from the membrane programs read from file.

\section{Model checking}  \label{sec:modelchecking}

	Model checking is an automated verification technique that explores all possible executions of a system to check whether it meets a given specification, involving different techniques and multiple variations. Model-checking models are traditionally based on annotated transition systems known as \emph{Kripke structures} $\mathcal K = (S, \to, I, AP, \ell)$ and consist of a set of states $S$, a binary relation $(\to) \subseteq S \times S$,\footnote{For simplicity, it is usually assumed that all executions of a Kripke structure are non-terminating, and so either $\to$ has a successor for every state, or finite executions are \emph{stutter-extended} by repeating their final state forever, like in Spin~\cite{spinmc} and other verification tools.} a finite set of initial states $I \subseteq S$, a finite set of atomic propositions $AP$, and a labeling function $\ell : S \to \mathcal P(AP)$ that associates them to each state. Properties are usually expressed in terms of these atomic propositions using some temporal logics endowed with operators to specify how they occur in time. Well-known examples are CTL*~\cite{ctlstar} and its sublogics LTL (Linear Temporal Logic)~\cite{pneuliLTL} and CTL (Computational Tree Logic)~\cite{ctl}, and $\mu$-calculus~\cite{mucalcmc}.

	Rewriting systems can be naturally seen as Kripke structures whose states are terms and whose transitions are one-step rule rewrites. For their part, the natural model for P systems has membrane configurations as states and evolution steps as transitions. Mapping membrane configurations to terms is straightforward, as we have seen in~\cref{sec:repr}, but matching evolution steps and rewrite rules is not. In fact, this is related to the general problem of describing parallel with sequential rewriting. However, strategies and the possibility to consider them as atomic transitions in our strategy-aware model checker will solve this problem.

	\paragraph{Strategy-aware model checking} Maude includes an on-the-fly LTL model checker~\cite{maudemc} since its 2.0 version, which we have recently extended to support systems controlled by strategies~\cite{fscd}. Looking at a strategy as a subset $E \subseteq S^* \cup S^\omega$ of allowed executions of a model $\mathcal K$, the satisfaction of a linear-time property $(\mathcal K, E) \vDash \varphi$ can be understood as its satisfaction $\mathcal K, \pi \vDash \varphi$ for all allowed executions $\pi \in E$. The question of which are the executions described by a Maude strategy language expression is given answer by a small-step operational semantics, which is respected by the model-checker implementations. This can be easily extrapolated to branching-time properties~\cite{btimemc}, whose allowed executions can be seen as a restricted execution tree. CTL, CTL*, and $\mu$-calculus properties can be checked using external model checkers through an extensible interface \textsf{umaudemc} that unifies the interaction and the syntax of the logics~\cite{umaudemc}. This interface is built over a library that allows accessing Maude objects and operations from Python and other programming languages, which can be used to make these model checkers directly available to Maude-based frameworks like the one for membrane systems presented here.

	\paragraph{Model preparation and atomic properties} In both the original and the strategy-aware Maude model checkers, users should specify the atomic propositions as regular operators of a predefined sort \texttt{Prop}, and its satisfaction relation by equations on a predefined symbol \texttt{\_|=\_}~\cite[\S 12]{maude}. For membrane systems, we provide a general set of predefined properties that include, among others:
\label{code:preds}
\begin{lstlisting}
mod P-SYSTEM-PREDS is
  protecting P-SYSTEM-CONFIGURATION .
  including SATISFACTION .
  protecting EXT-BOOL .

  subsort Soup < State .

  op isAlive  : MembraneName      -> Prop [ctor] .
  op contains : MembraneName Soup -> Prop [ctor] .

  op {_}      : BoolExpr          -> Prop [ctor] .
  op _=_      : NatExpr NatExpr   -> BoolExpr [ctor] .
  op _+_      : NatExpr NatExpr   -> NatExpr [ctor] .
  op count    : MembraneName Soup -> NatExpr [ctor] .
  
  *** [...]
endm
\end{lstlisting}
The property \texttt{isAlive} checks whether a membrane is present in the configuration, and \texttt{contains} whether it contains some objects. More complex properties can be built with Boolean and integer expressions of sorts \texttt{BoolExpr} and \texttt{NatExpr} between curly brackets. The multiplicities of any multiset in any membrane, designated with the \texttt{count} operator, can be combined with the arithmetical operators and relations supported by Maude. For example, the property \texttt{\{ count(M1, a) = 2 * count(M2, b) \}} says that the number of \texttt{a}s in \texttt{M1} doubles the number of \texttt{b}s in \texttt{M2}. The evaluation of these expressions and the satisfaction of these propositions is defined equationally in the same module.

	Finally, the model checker is accessed through a special \texttt{modelChecker} operator declared in the predefined \texttt{STRATEGY-MODEL-CHECKER} module.
\begin{lstlisting}[escapechar=^]
op modelCheck : State Formula Qid QidList Bool 
                ~> ModelCheckResult [special (^\ldots^)] .
\end{lstlisting}
Its arguments are the initial state, the LTL formula to be checked, and the name of a strategy to control the system, plus two other optional arguments. The fourth one is very useful for membrane systems: a list of named strategies whose executions must be considered as atomic steps of the verified system. Like this, the executions of an \texttt{mpr} step can be automatically seen as the steps of the model, making the \emph{next} operator of the temporal logics work as expected and hiding the intermediate states in which the rules that are supposed to be executed in parallel are being applied. Thus, issuing
\begin{lstlisting}[language={}]
red modelCheck(< M1 | a b < M2 | a > >, 
               [] contains(M1, a), 'mcomp, 'mpr) .
\end{lstlisting}
will check the property that \texttt{M1} always contains an \texttt{a} in all membrane executions from the given initial one. The interactive environment of the next section will do all this behind the scenes.

\section{The membrane system environment} \label{sec:environ}

	The executable rewriting logic framework proposed to represent membrane systems can be directly instantiated with the specification of a particular system in Maude itself: extending \texttt{P-SYSTEM-STRATEGY} with the declaration of its objects and its membranes, their evolution rules as rewrite rules, and their ascription to a membrane with strategies. However, dealing with priorities or other extensions is not so simple, and can be automatically done by convenient program transformations. The interactive environment described in this section implements these manipulations and allows simulating and verifying membrane systems easily. These should be specified in an extended version of the membrane description language of a previous prototype in~\cite{membrane}, on which ours was initially based.

	After downloading the membrane example from \href{http://maude.ucm.es/strategies}{\ttfamily maude.ucm.es/strategies}
and loading the \texttt{memrun.maude} file into Maude, the following command will execute the interactive environment:\footnote{In Maude 3.1, file operations are disabled by default for security reasons, so \texttt{-allow-files} must be given as a command line argument to Maude in order to use the environment.}

\begin{lstlisting}[language={}]
Maude> erewrite initREPL(repl) < repl : MemREPL | none > .

      ** Membrane system environment in Maude **

Membrane>
\end{lstlisting}
The environment offers different commands that are listed by typing \texttt{help}.
The \texttt{load} command reads a membrane specification from a file and runs the commands in it. For example, \texttt{load divisor.memb} loads the membrane system of~\cref{fig:venn}.
\begin{lstlisting}[language={}, escapechar=^]
Membrane> load divisors.memb
File ^{\color{olive}divisors.memb}^ has been loaded.
\end{lstlisting}
In that file, evolution rules (introduced by \kywd{ev}) and priorities (introduced by \kywd{pr}) are specified as shown below for the membrane \texttt{M2}:
\begin{lstlisting}[language=membrane]
membrane M2 is
  ev r21 : d a -> c .        ev r22 : c   -> d .
  ev r23 : tic -> tac .      ev r24 : a tac -> a tic .
  ev r25 : d tac -> d .      ev r26 : tac -> delta .
  pr r24 > r26 .
  pr r25 > r26 .
end
\end{lstlisting}
Loading the file implies generating the strategies described in~\cref{sec:repr} for later use by the various supported commands. They can be shown with the \texttt{show strats} command followed by the membrane name.
\begin{lstlisting}[language={}, moredelim={[is][\color{blue}]{\#}{\#}}]
Membrane> show strats M2 .
#Weak priority:# (r24 | r25 or-else r26) | r21 | r22 | r23
#Strong priority:# r21 ; mpr-strong(M2, ('r21, AR))
  | r22 ; mpr-strong(M2, ('r22, AR))
  | r23 ; mpr-strong(M2, ('r23, AR))
  | (r24 ; mpr-strong(M2, ('r24, AR))
    | r25 ; mpr-strong(M2, ('r25, AR))
  or-else match H s.t. intersection(('r24, 'r25), AR) =
          empty ; r26 ; mpr-strong(M2, ('r26, AR)))
\end{lstlisting}
If the command omits the \texttt{strats} word, the membrane definition is shown instead, and \texttt{show membranes} displays the names of all loaded membranes.

	The \texttt{trans} and \texttt{compute} commands allow simulating evolution steps and computations. The first one executes a single step, indicating the multiset of rules applied for each membrane.
\begin{lstlisting}[language={}, moredelim={[is][\color{blue}]{\#}{\#}}]
Membrane> trans < M1 | a a a tic < M2 | d tac > > .
Solution 1 with r11 r12 r13 in #M1#, r25 in #M2# :
	< M1 | c c c < M2 | a a a d d tic > >
Solution 2 with r12 r12 r12 r13 in #M1#, r25 in #M2# :
	< M1 | c c c < M2 | a a a d tic > >
No more solutions.
\end{lstlisting}
The \texttt{compute} command shows all irreducible states that can be found by successive transitions.
\begin{lstlisting}[language={}, moredelim={[is][\color{blue}]{\#}{\#}}]]
Membrane> compute < M1 | a a a a a a a a tic < M2 | empty > > .
Solution 1:	< M1 | d d d d >
Solution 2:	< M1 | < M2 | d d d > >
Solution 3:	< M1 | d d >
Solution 4:	< M1 | d >
No more solutions.
\end{lstlisting}
For this divisors calculator, the solutions tell us that 2 and 4 are the non-trivial divisors of 8, after reading their number of \texttt{d}s in \texttt{M1}.
The interpretation of rule priorities, either \texttt{weak} or \texttt{strong}, can be set globally with the \texttt{set priority} command that changes the definition of the \texttt{handleMembrane} strategy mentioned in~\cref{sec:repr}. By default, strong priorities are used.

	Temporal properties of the membrane executions can be checked with the \texttt{check} command. These properties are expressed in LTL for the predefined language of atomic propositions described in~\cref{code:preds}, which can anyhow be extended by modifying the environment source code. For example, the following command checks that the number of \texttt{d}s in the membrane \texttt{M1} is either 0 or a divisor of 12 in all reachable configurations from an initial membrane with 12 \texttt{a}s. Thus, a false divisor is never generated.
\begin{lstlisting}[language={}]
Membrane> check < M1 | a a a a a a a a a a a a tic
   < M2 | empty > > satisfies [] ({ count(M1, d) = 0 }
   \/ { count(M1, d) divides 12 }) .
The property is satisfied.
\end{lstlisting}
When the property is not satisfied, the output shows a counterexample describing the intermediate steps and the rules that have been applied. This is the case of the following property claiming that every state containing \texttt{tac} in \texttt{M2} is followed by a state containing \texttt{tic}.
\begin{lstlisting}[language={}, mathescape, escapechar=^, moredelim={[is][\color{blue}]{\#}{\#}}]
Membrane> check < M2 | a a d d tic >
  satisfies [] (contains(M2, tac) -> O contains(M2, tic)) .
^\color{red}|^ < M2 | a a d d tic >
$\color{red}\kern-.8pt\vee$ with r21 r21 r23 in #M2#
^\color{red}|^ < M2 | c c tac >
$\color{red}\kern-.8pt\vee$ with r22 r22 r26 in #M2#
^\color{green}X^ < M2 | delta d d >
\end{lstlisting}

The \texttt{check} command admits bounded model checking on the number of objects in the configuration, where the bound may be indicated between brackets after the \texttt{check} keyword. This is useful for membrane systems that, unlike this example, have an unbounded configuration space. For instance, the following membrane system calculates the squares $\langle M_1 \mid d^n \, e^{n^2} \rangle$ of all natural numbers $n \geq 1$ starting from $\langle M_1 \mid \langle M_2 \mid \langle M_3 \mid a \, f \rangle \,\! \rangle \,\! \rangle$.
\begin{lstlisting}[language=membrane]
membrane M1 is end

membrane M2 is
  ev r21 : b   -> d .      ev r22 : d   -> d e .
  ev r23 : f f -> f .      ev r24 : f   -> delta .

  pr r23 > r24 .
end

membrane M3 is
  ev r31 : a -> a b .
  ev r32 : a -> b delta .
  ev r33 : f -> f f .
end
\end{lstlisting}
The membrane \texttt{M3} nondeterministically produces a number $n \geq 1$ of \texttt{b}s along with $2^n$ \texttt{f}s in $n$ steps, and then spills its contents into \texttt{M2}. Then \texttt{M2} generates one \texttt{e} for each \texttt{d} in every one of the $n$ steps required to reduce the exponential number of \texttt{f}s with \texttt{r23}, hence calculating $n^2$.
\begin{lstlisting}[language={}, moredelim={[is][\color{olive}]{\#}{\#}}]
Membrane> load nsquare.memb 
File #nsquare.memb# has been loaded.
Membrane> compute [3] < M1 | < M2 | < M3 | a f > > > .
Solution 1:	< M1 | d e >
Solution 2:	< M1 | d d e e e e >
Solution 3:	< M1 | d d d e e e e e e e e e >
No more solutions requested.
\end{lstlisting}
The (not so) atomic proposition \lstinline@{ count(M1, d) ^ 2 = count(M1, e) }@ claims that the number of \texttt{e}s is the square of the number of \texttt{d}s in \texttt{M1}. This property cannot be checked as an invariant on the whole infinite state space, but it can be, for instance, in the reachable portion of the model where the number of objects is always below 70.
\begin{lstlisting}[language={}]
Membranes> check [70] < M1 | < M2 | < M3 | a f > > >
  satisfies [] { count(M1, d) ^ 2 = count(M1, e) } .
The property is satisfied.
\end{lstlisting}

	Moreover, membrane systems can also be checked against CTL* and $\mu$-calculus properties, using the external model checkers for strategy-controlled systems~\cite{btimemc}. Since these are not integrated in the Maude interpreter, they should be used through an external tool instead of the membrane environment. For instance, resuming the divisor calculator example, we can check the $\mu$-calculus property $\nu Z . \, \mathit{isAlive}(M_2) \wedge [\cdot] (\neg \, \mathit{isAlive}(M_2) \vee [\cdot] \, Z)$ asserting that the membrane $M_2$ is only dissolved in odd steps.
\begin{lstlisting}[language={}]
$ ./membranes.py -v divisors.memb \
   '< M1 | a a a a a a a a a a a a tic < M2 | empty > >' \
   'nu Z . (isAlive(M2) /\ [.] (~ isAlive(M2) \/ [.] Z))'
Rewriting model generated in 4101 rewrites.
The property is satisfied (70 states, 76884 rewrites).
\end{lstlisting}
In fact, the first clause $\mathit{isAlive}(M_2)$ claims that $M_2$ is present in the configuration, and the second one requires that after any possible step $\neg \, \mathit{isAlive}(M_2) \vee [\cdot] \, Z$ is satisfied, meaning that either $M_2$ is no longer present or in the next step the property holds again, as mandated by the fixed point $\nu$. This property cannot be expressed in CTL*. Checking whether at least a divisor of 12 is found can be done with the following CTL property, although this is also possible by simulating with the \texttt{compute} command.
\begin{lstlisting}[language={}]
$ ./membranes.py divisors.memb \
   '< M1 | a a a a a a a a a a a a tic < M2 | empty > >' \
   'E <> { count(M1, d) divides 12 }'
The property is satisfied (70 system states, 91136 rewrites).
\end{lstlisting}
Model checking in a bounded subset of the state space is also possible with this interface, both on the number of objects and on the number of evolution steps.

\subsection{Implementation notes}

	This prototype is based on the Maude specification of membrane systems in~\cref{sec:repr}, and uses Maude reflection to construct the rules and strategies of the specific membrane systems parsed from the input. The external objects introduced in Maude~3 are used to read commands and write their results to the terminal, and also to read membrane specifications from external files. In either case, text is read as a string, and parsed using the \texttt{metaParse} descent function in specific modules describing the grammar of the different elements.

	Evolution rules are directly translated to rewrite rules, but loose objects in the right-hand side are enclosed into a \texttt{here} target, because targets are used to exclude consumed terms from being rewritten twice in the same evolution step. For some commands, the right-hand side of rules is appended a \texttt{log} object with its label to get track of the rules that have been applied, without interfering with the execution of the system. This allows showing them in the \texttt{trans} command and the model-checking counterexamples.

	For example, the following equation is the actual translation from parsed evolution rules (containing unparsed fragments, known as bubbles) in the membrane specification to Maude rewrite rules at the metalevel.
\begin{lstlisting}[moredelim={[is][]{\#}{\#}}]
ceq makeRules(M, 'ev_:_->_.['token[Q], 'bubble[LHS],
                                'bubble[RHS]]) =
   (#rl# PLHS => PRHS [label (downTerm(Q, 'UNNAMED))] .)
  if PLHS := getTerm(metaParse(M, none,
               downTerm(LHS, (nil).QidList), 'ObjSoup))
  /\ T    := getTerm(metaParse(M, none,
               downTerm(RHS, (nil).QidList), 'Soup))
  /\ PRHS := getTerm(metaReduce(M, 'wrapHere[T])) .
\end{lstlisting}
The module \texttt{M} would be an extension of \texttt{P-SYSTEM-CONFIGURATION} (see~\cref{sec:repr}) populated with the inferred signature of objects for the particular membrane system. The left-hand side is parsed in the \texttt{ObjSoup} sort of this module, while the right-hand side is a \texttt{Soup} allowed to contain targets. The function \texttt{wrapHere} wraps the free objects of the right-hand side into a \texttt{here} target, as explained before.

	After the membrane specification is parsed, strategies are generated for the prioritized versions of the maximal parallel step. For instance, the procedure to generate the weak priority strategy explained in~\cref{sec:repr} is executed by a fixed-point equational computation that starts with the minimal elements of the relation
\begin{lstlisting}
op genPriorityStrat : QidSet PriorityRelation -> Strategy .
op genPriorityStrat : Strategy PriorityRelation -> Strategy .

eq genPriorityStrat(Rs, PR) =
  genPriorityStrat(genRuleApps(minimal(Rs, PR)), PR) .

op genRuleApps : QidSet -> Strategy .
eq genRuleApps(none) = fail .
eq genRuleApps(R ; Rs) = (R[none]{empty}) | genRuleApps(Rs) .
\end{lstlisting}
and iterates extending the newly introduced rules until the strategy is stabilized.
\begin{lstlisting}[escapechar=^, moredelim={[is][]{\#}{\#}}]
eq genPriorityStrat(S, PR) =
  #if# orElseSimplify(extendPrec(S, PR)) == S
  then S else genPriorityStrat(
                orElseSimplify(extendPrec(S, PR)), PR) fi .

op extendPrec : Strategy PriorityRelation -> Strategy .

eq extendPrec(S1 or-else S2, PR) =
     extendPrec(S1, PR) or-else S2 .
ceq extendPrec(S1 | S2, PR) = extendPrec(S1, PR) 
                            | extendPrec(S2, PR)
 if S1 =/= fail /\ S2 =/= fail .
eq extendPrec(R[none]{empty}, PR) = #if# pred(R, PR) == none
  then R[none]{empty}
  else genRuleApps(pred(R, PR)) or-else R[none]{empty} fi .
\end{lstlisting}
where \texttt{pred} returns the predecessors of a rule in the priority relation \texttt{PR}. The strategy for strong priorities is generated similarly. Reflective module transformations assign the adequate \texttt{handleMembrane} definition depending on how rule priorities are understood.

The command-line utility to model check against branching-time properties generates the membrane system theory using the same equational infrastructure of the interactive environment. However, instead of using Maude external objects for reading files and printing messages, the standard Python library is used, in which it is programmed. Once the model is set up, it is a strategy-controlled Maude specification that is directly passed to the unified model-checking tool \textsf{umaudemc}~\cite{btimemc} via its Python API.

\section{Variations of membrane systems} \label{sec:variations}

	Many variants of membrane systems have been proposed to better address different applications and problems~\cite{membraneComputingIntro,membraneApplications}. The flexibility of the Maude language and its strategies makes modifying the prototype to support and experiment with these variations a relatively easy task. In this section, we illustrate this extensibility with three widespread features. Each subsection starts by describing and motivating one of the variants, then explains how it is implemented in the prototype, and finally shows the feature in action with an example. In \cref{sec:division}, we also discuss how to fix the order in which membranes are processed to avoid redundant calculations, as anticipated in the previous sections.

\subsection{Structured objects}

	Standard membrane systems operate on multisets of unstructured opaque objects, but chemicals in a cell are usually complex molecules (DNA, proteins,~\ldots) which are better described by structured data like strings or trees. \emph{String rewriting P systems}~\cite{membraneString,membraneApplications} are membrane systems whose objects are strings made out of terminal and nonterminal symbols, and whose evolution rules have the form $A \to w$ where $A$ is a nonterminal symbol and $w$ is a word. Targets are expressed similarly. Evolution rules are applied like in the standard P systems, but only one rule can be applied to each string in the membrane soup at each evolution step.

	A generalization of this possibility has been implemented in the rewriting logic framework presented here, where the language of objects of the membrane system consisted of some plain identifiers, implicitly declared as they are used in the membrane specification. In this section, the user will be allowed to explicitly declare the signature of objects, which may include arbitrary Maude terms in their arguments to be considered modulo equations and axioms. Specific syntax in the membrane specification language is devoted to the definition of objects. For instance, the following lines declare string objects on some constants \texttt{a}, \texttt{b}, and \texttt{c}. The associative binary operator \texttt{\_·\_} stands for string concatenation, which is associative, and \texttt{eps} is the empty word.
\begin{lstlisting}[language=membrane, mathescape]
signature is
  ob _$\cdot$_ : Obj Obj [assoc id: eps prec 30] .
  obs a b c eps .
end
\end{lstlisting}
Object-declaration statements are similar to regular Maude operator declarations except that the range sort is omitted and that they are introduced by the \kywd{ob} or \kywd{obs} keywords. The sort \texttt{Obj} of objects is the implicit range of all object declarations. Maude functional modules may be imported with the \texttt{import} statement, as shown in the example of the following section. Orthodox string rewriting P systems can be defined on top of this signature by using only evolution rules of the form $A \to u$, but different signatures and other types of rules can be tried instead.

	The technical difficulty of applying rules in structured objects is ensuring that each object is only rewritten once in each step, as required for strings. Moreover, if evolution rules were translated as in the basic model, they could get applied on the arguments of structured objects with undesired results, like targets appearing inside objects. Hence, the different types of rules have to be considered carefully. Rules $r : u \to (v_1, m_1) \cdots (v_n, m_n)$ with multiple targets $m_k$ can only be applied at the top multiset, so they are executed with the strategy
\begin{lstlisting}[mathescape]
matchrew O R by O using top($r$)
\end{lstlisting}
where \texttt{O} and \texttt{R} are variables of sorts \texttt{Obj} and \texttt{Soup}. This precaution is unnecessary if $u$ is a multiset of objects or a single object that would never appear in a nested context, so this wrapper can be avoided in these cases. Any rule with a single target $r : t \to (t', m)$ can in principle be applied anywhere, either at the top multiset or at any position $p$ of an object $o$. If $t$ matches $o$ in $p$ with substitution $\sigma$, then the object $o$ must become the target $(o[p/\sigma(t')], m)$. This is achieved by transforming the evolution rule $r$ to the rewriting rule $r' : t \to t'$, and wrapping the object with the appropriate target once rewritten using an auxiliary rule \texttt{wrapMsg}.
\begin{lstlisting}[mathescape]
matchrew O R by O using $r'$ ; wrapMsg[M <- $m$]
rl [wrapMsg] : R => (R, M) [nonexec] .
\end{lstlisting}

	Extending the membrane specification language, we add the declaration statement \kywd{xev} for those rules that are meant to be applied inside objects, which can have at most one target.
\begin{lstlisting}[language=membrane, mathescape]
xev $\mathit{lbl}$ : $t$ -> ($t'$, $m$) .
\end{lstlisting}
Rules declared with \kywd{ev} are applied only at the top even if they can match inside an object. Rules whose left-hand side is a multiset are not wrapped for efficiency, because object multisets are assumed to only appear at the top level.

	The following simple membrane system defines three evolution rules on the object string signature presented before.
\begin{lstlisting}[language=membrane, mathescape]
membrane M1 is
  xev s1 : a $\cdot$ a -> a .
  xev s2 : b -> c $\cdot$ c .
  ev  s3 : a -> c .
end
\end{lstlisting}
The rule \texttt{s1} removes duplicated \texttt{a}s in strings, while \texttt{s2} transforms each \texttt{b} in a pair of \texttt{c}s. The last rule \texttt{s3} transforms loose \texttt{a} objects into \texttt{c}, but it is not applied on the characters of a string because it uses the \kywd{ev} instead of the \kywd{xev} keyword.
\begin{lstlisting}[language={}, escapechar=^, mathescape, moredelim={[is][\color{blue}]{\#}{\#}}]
Membrane> load strings.memb
File ^{\color{olive}strings.memb}^ has been loaded.
Membrane> trans < M1 | (a $\cdot$ a $\cdot$ b $\cdot$ a) b (a $\cdot$ a) > .
Solution 1 with s1 s1 s2 in #M1# :
	< M1 | (a $\cdot$ b $\cdot$ a) (c $\cdot$ c) a >
Solution 2 with s1 s2 s2 in #M1# :
	< M1 | (a $\cdot$ a $\cdot$ c $\cdot$ c $\cdot$ a) (c $\cdot$ c) a >
No more solutions.
Membrane> compute < M1 | (a $\cdot$ a $\cdot$ b $\cdot$ a) b (a $\cdot$ a) > .
Solution 1:   < M1 | (a $\cdot$ c $\cdot$ c $\cdot$ a)  (c $\cdot$ c) c >
No more solutions.
\end{lstlisting}
The \texttt{show strats} command lets us observe that the \texttt{membraneRules} strategy is generated as explained above.
\begin{lstlisting}[language={}, moredelim={[is][\color{blue}]{\#}{\#}}]
Membrane> show strats M1 .
#Weak priority:#  matchrew O R by O using top(s3) 
  | matchrew O R by O using(s1 ; top(wrapMsg[T <- here]))
  | matchrew O R by O using(s2 ; top(wrapMsg[T <- here]))
\end{lstlisting}
Another example with structured objects, but without subterm rewriting, is shown in the following section.

\subsection{Membrane division} \label{sec:division}

	Another important and popular extension of membrane systems is membrane division, inspired on the cellular \emph{mitosis}, that allows membrane systems to grow and take advantage of the parallelism of its computation model. In our realization, membrane division is triggered by a new target that replaces the affected cell by two copies of itself with all of its contents, plus a different set of objects for each copy included in the target triple. The following declaration and rule should be added to \texttt{P-SYSTEM-CONFIGURATION}:
\begin{lstlisting}
op `(_,_,div`) : ObjSoup ObjSoup 
                 -> TargetMsg [ctor frozen (1 2)] .

rl [div] : < MN' | EW < MN | CW (W, W', div) > >
        => < MN' | EW < MN | CW W > < MN | CW W' > > .
\end{lstlisting}
Division is done by the new rule \texttt{div} above, which prevents duplicating the outermost membrane. This rule should be applied exhaustively in the third phase of the evolution steps, just after object communication has been completed, but before membranes are dissolved. This involves changing the \texttt{mpr} strategy definition:
\begin{lstlisting}
sd mpr := visit-mpr ; amatch TM ;
          communication ; (div !) ; (dis !) .
\end{lstlisting}
Other cell operations like creation, merge, endocytosis (introducing a membrane into another one), exocytosis (expelling a membrane), and gemmation could be implemented similarly.

	Combining both membrane division and structured objects, the following membrane system implements a Boolean satisfiability (SAT) solver that runs in a polynomial number of evolution steps on the size of the formula. Each logical variable triggers a membrane duplication where each copy evaluates a possible value of the variable, making the membrane grow exponentially to evaluate in parallel all possible valuations. Formulae are specified using acyclic graphs indexed by natural numbers from the Maude's \texttt{NAT} module, with $0$ being the root of the formula. Each node is given by a symbol whose first argument is its own identifier, followed by the values or identifiers of the node arguments. The role of the \texttt{splitoken} object will be explained later.
\begin{lstlisting}[language=membrane, escapechar=^, moredelim={[is][]{\#}{\#}}]
signature is
  import NAT .

  ob  const  : Nat Bool . ^\hfill^ *** Logical constant
  ob  #var#    : Nat . ^\hfill^ *** Variable
  ob  not    : Nat Nat . ^\hfill^ *** Negation
  obs and or : Nat Nat Nat . ^\hfill^ *** Binary operators

  ob splitoken . ^\hfill^ *** Token to limit splitting
end
\end{lstlisting}
For example, $x \wedge \neg\, x$ can be written \texttt{and(0, 1, 2) var(1) not(2, 1)}. The skin membrane \texttt{M1} is the output membrane of the SAT solver, where the presence of an object \texttt{const(0, true)} indicates the satisfaction of the formula. However, we define it empty \lstinline[language=membrane, basicstyle=\ttfamily]{membrane M1 is end} as a mere receptor of the objects from \texttt{M2}. The \texttt{M2} membrane includes several rules to simplify the expressions, a rule \texttt{split} to fork the membrane with two copies where a variable takes alternatively the \texttt{true} and \texttt{false} values, and a rule \texttt{end} that dissolves the membrane when the evaluation has finished. Variables can be declared with the \kywd{var} statement as in Maude and used in the evolution rules. In this case, they match the integer indices and Boolean constants in the nodes.
\begin{lstlisting}[language=membrane, moredelim={[is][]{\#}{\#}}]
membrane M2 is
  var H M N : Nat .
  var B     : Bool .

  ev split : #var#(H) splitoken -> splitoken
               (const(H, true), const(H, false), div) .

  ev not   : not(H, N) const(N, B) 
          -> const(H, #not# B) const(N, B) .
  ev and1  : and(H, M, N) const(M, false)
          -> const(H, false) const(M, false) .
  ev and2  : and(H, M, N) const(N, false)
          -> const(H, false) const(N, false) .
  ev and3  : and(H, M, N) const(M, true) const(N, true) 
          -> const(H, true) const(M, true) const(N, true) .
  ev or1   : or(H, M, N) const(M, true) 
          -> const(H, true) const(M, true) .
  ev or2   : or(H, M, N) const(N, true) 
          -> const(H, true) const(N, true) .
  ev or3   : or(H, M, N) const(M, false) const(N, false)
        -> const(H, false) const(M, false) const(N, false) .

  ev #end#   : const(0, B) -> const(0, B) delta .

  pr not and1 and2 and3 or1 or2 or3 #end# > split .
end
\end{lstlisting}
Note that the \texttt{split} rule requires the object \texttt{splitoken} to be applied. This way, the number of cell divisions in each evolution step is limited to the number of such tokens, and some unnecessary cell divisions can potentially be avoided thanks to the simplification rules applied in the meanwhile.

The specification of this membrane system can be improved in many ways. For example, the \texttt{and} and \texttt{or} connectives are commutative, and so they can be described using a Maude-defined commutative pair that would reduce the number of rules. Moreover, each constant can only be used once in each evolution step as they are consumed by the evolution rule. This can be improved using promoters, which are introduced in~\cref{sec:promoters}. Finally, and more importantly, the election of the next variable to be assigned is nondeterministic. Since the order in which variables are assigned does not affect the satisfaction of the formula, this unnecessarily and exponentially increases the size of the state space of the strategic execution. The depth-first variant of the \texttt{compute} command \texttt{dfs compute} is more convenient for evaluating this system than the default breadth-first search variant.

	As advanced in \cref{sec:repr}, the strategy in charge of visiting the nested membranes of the configuration structure, \texttt{nested-mpr}, evaluates them in all possible orders, since all possible matches of the set-like argument will be tried at each call. These orders are factorially-exponentially many in the last example while the order in which membranes are processed is irrelevant (at least for the classes of membrane systems here considered). Hence, we must avoid it by fixing an order on the membranes, even at the expense of clarity.
\begin{lstlisting}[moredelim={[is][]{\#}{\#}}]
op orderMembranes : MembraneSoup -> MembraneList .
eq orderMembrane(empty) = nil .
eq orderMembrane(M MS) = insert(orderMembrane(MS), M) .

op insert : MembraneList Membrane -> MembraneList .
eq insert(nil, M) = M .
eq insert(M1 ++ ML, M) = #if# lt(M, M1) then M ++ M1 ++ ML 
                         else M1 ++ insert(ML, M) fi .
\end{lstlisting}
Using the \texttt{orderMembranes} function, \texttt{nested-mpr} can take a list of membranes (here assembled with the \texttt{++} symbol) instead of a set, and so process the membranes in a single fixed order. Membranes are ordered here using the \texttt{lt} operator of the \texttt{TERM-ORDER} module included in the Maude distribution that allows comparing arbitrary terms. However, this sorting algorithm is not really efficient and it is executed each time a membrane is processed. Sacrificing the due confluence modulo axioms of equational specifications, \texttt{orderMembrane} can be defined by the sole equation below, which will operationally fix the order in which the implementation visits the set.
\begin{lstlisting}
eq orderMembrane(M MS) = M ++ orderMembranes(MS) .
\end{lstlisting}
Unfortunately, while this situation is frequent and the Maude strategy language includes an operator \skywd{one} that stops when the first solution is found, there is no builtin support for stopping at the first match in strategy calls or \texttt{matchrew}s.

	For example, the SAT membrane system can be used to check the formulae $x \wedge \neg\, x$ and $(x_1 \vee x_2) \wedge (\neg \, x_1 \vee \neg\, x_2)$. The \texttt{const(0, true)} will not be found in the solution of the first one because it is unsatisfiable, but it will be for the second formula.
\begin{lstlisting}[language={}, escapechar=^, moredelim={[is][\color{olive}]{\#}{\#}}]
Membrane> load sat.memb
File #sat.memb# has been loaded.
Membrane> compute < M1 | < M2 | splitoken and(0, 1, 2)
                                var(1) not(2, 1) > > .
Solution 1:	< M1 | const(0, false) ^\ldots^ >
No more solutions.
Membrane> dfs compute [1] < M1 | < M2 | splitoken
	and(0, 5, 6) var(1) var(2) not(3, 1) not(4, 2)
	or(5, 1, 2) or(6, 3, 4) > > .
Solution 1:     < M1 | const(0, true) ^\ldots^ >
No more solutions.
\end{lstlisting}
The second formula is evaluated quickly but not immediately with \texttt{compute}, so \texttt{dfs compute [1]} is used to find the first solution by depth. Additional evolution rules could be defined in \texttt{M1} and \texttt{M2} to clean the irrelevant objects and to allow recovering the valuation.

\subsection{Promoters and inhibitors} \label{sec:promoters}

	Standard evolution rules only depend on the objects appearing in their left-hand side, which are consumed in its application. However, inspired in biochemical reactions, some processes may only take place in the presence of certain objects that are not part of the reaction. Conversely, some objects may act as inhibitors that impede a reaction to take place. This leads to rules with \emph{promoters} and \emph{inhibitors}~\cite{membranePromoters,membraneApplications}. The general form of these rules is $u \to v \mid_z$ and $u \to v \mid_{\neg z}$, meaning that the usual evolution rule $u \to v$ can only be applied if the objects $z$ distinct from $u$ are present (respectively not present) in the membrane. The objects $z$ are not consumed and can be used in the same evolution step.

	Promoters and inhibitors have already been considered in rewriting logic and Maude~\cite{memrewlogPromoters}. The authors distinguish two semantics depending on whether promoters and inhibitors are allocated statically or dynamically. Static allocation corresponds to the description of this feature in the previous paragraph, and to the theoretical simultaneity of rule application. Dynamic allocation is more easily expressed in rewriting logic, where evolution rules are applied sequentially, since it checks the presence of promoters and inhibitors on the intermediate state when some rules have already been applied and consumed part of the membrane contents. Their executable implementation in Maude only supports the latter, but we will implement the more widespread static allocation.

	Given a rule $u \to v$ with promoters $p$ and inhibitors $h$, we will generate a conditional rewriting rule of the form
\begin{lstlisting}[mathescape, escapechar=^]
crl $u$ => $v$ if $u$ $p$ SRest := S0
           /\ ^not^ contains($u$ $h$, S0) [nonexec] .
\end{lstlisting}
where the free variable \texttt{S0} occurs. This variable will be instantiated by providing an initial substitution to the rule application strategy with the contents of the membrane where the rule is applied at the beginning of the evolution step. Promoters are handled by a matching condition \texttt{:=} that matches into the initial membrane multiset a soup containing both the rule left-hand side, the promoters, and possibly something else. In case the promoter objects contain free variables, these will be instantiated and the rule will be executed for all their possible matches. Inhibitors cannot bind new variables, being objects that are not present in the configuration, so they are handled by the equationally-defined \texttt{contains} predicate. This function decides whether its first argument is contained in the second argument, i.e.\ whether both the rule left-hand side and the inhibitors are included in the initial contents.
A more strategic solution would have defined $u \to v$ directly and preceded its application by the tests \lstinline[mathescape, moredelim={[is][]{\#}{\#}}]|match W s.t. contains($u$ $p$, W0) /\ #not# contains($u$ $h$, W0)|, but $u$ may contain variables that must take the same values in the test and the rule application, thus complicating the correct definition of the strategy. Changes need also be made to strategies, which should pass the initial membrane contents to the evolution rule applications.
\begin{lstlisting}[mathescape, moredelim={[is][\color{red}]{\#}{\#}}]
sd visit-mpr := matchrew < MN | S MS > by S
  using handleMembrane(MN, #S#), MS using nested-mpr(MS) .
sd handleMembrane(MN, #S0#) := inner-mpr(MN, #S0#) .
sd inner-mpr(MN, #S0#) := (matchrew S TS by S 
                           using membraneRules(MN, #S0#)) ! .
sd membraneRules($M_i$, #S0#) := $r_1$ | $\ldots$ | $r_n$ 
                          | $r_1'$[S0 <- #S0#] | $r_m'$[S0 <- #S0#] .
\end{lstlisting}
The initial multiset is matched by the \texttt{S} variable of the \texttt{matchrew} operator in the \texttt{visit-mpr} strategy, and then, it is passed through strategy arguments to the definitions where rules with promoters and inhibitors are applied using the \texttt{[S0 <- S0]} initial substitution.
Similar modifications are suffered by the strategies implementing the prioritized application of rules.

	In order to allow expressing rules with promoters and inhibitors, the membrane specification language has been extended: these rules start with the \kywd{cev} keyword and finish with the specification of those multisets after the \kywd{with} or \kywd{without} keyword (both or only one can be used). The multisets $p$ and $h$ may contain any variable that occurs in the left-hand side of the rule.
\begin{lstlisting}[language=membrane, mathescape]
cev $\mathit{lbl}$ : $u$ -> $v$ with $p$ without $h$ .
\end{lstlisting}
Moreover, this syntax and the transformation described before can be easily extended to support more complex conditions or guards for evolution rules, like those used in \emph{kernel P systems}~\cite{kernelPSystem} that include integer expressions on the multiplicity of arbitrary subsets in the initial multiset, in which promoters and inhibitors can be expressed.

	Using this new feature, the SAT solver system of the previous section can be simplified and made more efficient. While the signature and the overall structure of the rules is kept unchanged, the common terms in both sides of evolution rules are placed as promoters, so that they can be used more than once in each step. In addition, the rule \texttt{split} on the variable \texttt{H} is inhibited by the object \texttt{var(s(H))}, hence forcing the variables to be assigned in decreasing index order (assuming they are numbered consecutively), and so limiting nondeterminism.
\begin{lstlisting}[language=membrane, moredelim={[is][]{\#}{\#}}]
membrane M1 is end

membrane M2 is
  var H M N : Nat .
  var B     : Bool .

  cev split : #var#(H) splitoken -> splitoken
              (const(H, true), const(H, false), div)
              without #var#(s(H)) .

  cev not  : not(H, N) -> const(H, #not# B) with const(N, B) .
  cev and1 : and(H, M, N) -> const(H, false)
        with const(M, false) .
  cev and2 : and(H, M, N) -> const(H, false) 
        with const(N, false) .
  cev and3 : and(H, M, N) -> const(H, true
        with const(M, true) const(N, true) .
  cev or1  : or(H, M, N)  -> const(H, true)
        with const(M, true) .
  cev or2  : or(H, M, N)  -> const(H, true)
        with const(N, true) .
  cev or3  : or(H, M, N)  -> const(H, false) 
        with const(M, false) const(N, false) .

  ev #end#   : const(0, B) -> const(0, B) delta .

  pr not and1 and2 and3 or1 or2 or3 #end# > split .
end
\end{lstlisting}
This membrane specification can be executed to check that the propositional formula $(x_1 \vee \neg\, x_3) \wedge (\neg\, x_2 \vee x_3 \vee \neg\, x_4)$ is satisfiable:
\begin{lstlisting}[language={}, escapechar=^, moredelim={[is][\color{olive}]{\#}{\#}}]
Membrane> load sat_promoters.memb
File #sat_promoters.memb# has been loaded.
Membrane> dfs compute [1] < M1 | < M2 | splitoken
                 var(1) var(2) var(3) var(4)
                 not(5, 3) not(6, 2) not(7, 4)
                 or(8, 1, 5) or(9, 6, 3) or(10, 9, 7)
                 and(0, 8, 10) > > .

Solution 1: < M1 | const(0, true) ^\ldots^ >
\end{lstlisting}
It can be observed using the \texttt{trans} command that the evolution of the membrane system is deterministic, finishes in 8 evolution steps, and uses up to 16 simultaneous \texttt{M2} membranes. The whole execution can be seen with the \texttt{check} command and the \verb|[] ~ contains(M1, const(0, true))| property.

\section{Performance considerations} \label{sec:performance}

	In this strategy-controlled rewriting framework for membrane systems, evolution steps are executed by exhaustively applying rules on each membrane, freely or according to some priorities. This involves visiting potentially many intermediate states for every admissible multiset of the membrane rules. Although computing maximal parallel steps cannot completely avoid this, more efficient algorithms are feasible by better planning the use of objects and compatible rules. For example, if the left-hand sides of two rules are disjoint multisets, all their interleaved applications will produce the same result, and so one can be executed exhaustively before the other.

	We have compared other Maude-based simulators and model checkers for membrane systems with ours. First, the prototype of the work \emph{Strategy-based proof calculus for membrane systems} by O. Andrei, and D. Lucanu~\cite{membrane} has been adapted to work with the current version of Full Maude, and the examples included in its distribution have been executed several times and measured with both prototypes. The only available command at their prototype is \texttt{trans}, and there are some bugs in its implementation that do not affect our comparison but prevent us from testing it with the other examples in this paper. On average, the new prototype is 9.47 times faster than the older, or 8.8 times if the initialization time of Maude and the prototypes is subtracted. \Cref{table:comparison} shows their execution times and their quotients for each example.

\begin{table}[h]\centering
\begin{tabular}{l r r r r r r}
	\toprule
				& \multicolumn6c{Time (ms)} \\
	\cmidrule{2-7}
	Prototype 		& ex1	& ex2	& ex3 	& ex4	& divisors	& nsquare 	\\
	\midrule
	SPCMS \cite{membrane}	& 897 	& 824	& 1014 	& 851	& 		&		\\
	ESPS \cite{exmcPsystem}	& 	&	&	&	& 1786		& 2945		\\
	This one		& 119   & 45	& 235	& 72	& 166		& 124		\\
	\midrule
	Speed-up		& 9.45	& 9.07	& 10.45	 & 8.9  & 10.76	 	& 26.47		\\
	Without init		& 9.12	& 5.14	& 15.65	 & 5.29 & 20.76		& 77.03		\\
	\bottomrule
\end{tabular}
\caption{Comparison with previous Maude-based prototypes.} \label{table:comparison}
\end{table}

	Moreover, our prototype has also been compared with that of the work \emph{Executable Specifications of P Systems} by O. Andrei, G. Ciobanu, and D. Lucanu~\cite{exmcPsystem} with support for model checking. In this case, the \texttt{divisors} and \texttt{nsquare} examples have been checked against the mentioned properties discussed in~\cref{sec:environ}. The divisor calculator was translated to the language of their prototype, and the square number generator is the example included in their distribution. This latter example is model checked in a bounded state space, which was fixed to 15 objects in their example and to 70 in~\cref{sec:environ}. Since a limit of 70 takes much time in their prototype (it has been canceled after 5 minutes, while ours finishes in 2.5 seconds), a bound of 35 objects was fixed for both.
	
\begin{figure}\centering
		\hbox{\kern-1em\includegraphics[scale=.4, page=1]{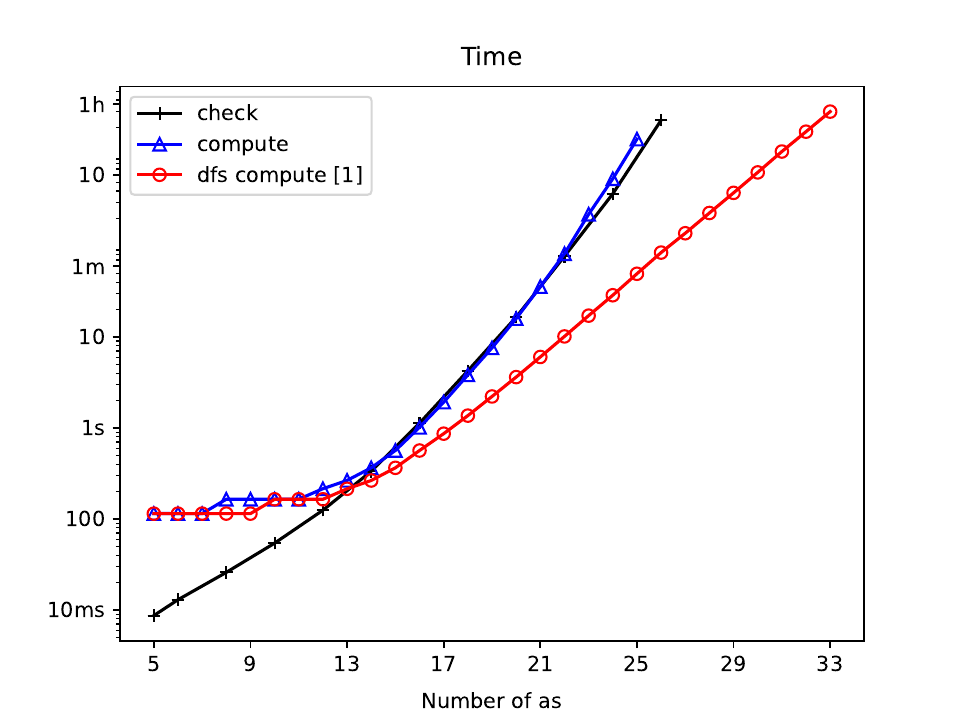} \kern-1em
		\includegraphics[scale=.4, page=2]{img/membrane.pdf}}
		\caption{Execution time and memory usage for different commands on the divisor calculator.} \label{fig:divstime}
	\end{figure}

	Nevertheless, if only the limit of 70 objects is increased to 71 in the previous property, the \texttt{check} command in our prototype does not finish within  an hour. Similarly, we have pushed the capabilities of our prototype to the limit with the other examples in this paper. For the \texttt{divisors} calculator, we have checked the $\mu$-calculus property in~\cref{sec:environ}, computed all irreducible configurations with \texttt{compute}, and a single solution by depth with \texttt{dfs compute} from the initial term $\langle M_1 \mid a^n \, \mathit{tic} \; \langle M_2 \mid \; \rangle \rangle$ on increasing $n$. As shown in~\cref{fig:divstime}, the execution time and the memory usage grow exponentially.\footnote{\Cref{fig:divstime} shows as \texttt{check} the execution time and memory usage of the \texttt{membranes.py} script using the Python-based builtin model checker (see~\cite{btimemc}). Since the measure of the builtin backend does not include the interface initialization time, their results for reduced sizes are notably smaller.} The only exception is the constant memory usage of the depth-first search \texttt{compute} command. Within an hour, results are obtained by \texttt{compute} and \texttt{check} for $n \leq 25$, and by the depth-first search of a single solution for $n \leq 33$. However, the results for $n \leq 23$ and $n \leq 28$ respectively have already been obtained in five minutes. For the SAT solver with promoters and inhibitors in~\cref{sec:promoters}, the depth-first search can stand up to 15 distinct variables in 5 minutes and to 17 in an hour.

	Regarding the different extensions in~\cref{sec:variations}, the performance penalty caused by them respect to the functionality of the basic prototype is relatively small, since these features only produce an additional cost when they are effectively used.
As mentioned in~\cref{sec:promoters}, we cannot compare the known implementation of promoters and inhibitors in Maude~\cite{memrewlogPromoters} with ours, since they use a different \emph{dynamic} interpretation of this feature.
We have also tried to compare the prototype with the k\textsc{PWorkbench} tool~\cite{modelCheckingMembranes} based on kernel P systems. However, the graph-like structure of these systems is not directly translatable to the nested structure of those used in this paper, its simulator randomly chooses an alternative for each evolution step while our command considers all of them, and the concepts of model used for model checking apparently do not coincide.

\section{Conclusions and future work}

	In this work, a rewriting logic framework controlled by rewriting strategies is proposed to express, simulate, and verify membrane systems. Strategies are used to bridge the gap between sequential rule rewriting and the parallel evolution of these systems by describing their particular control mechanisms. This approach is implemented as a metalanguage tool in the Maude specification language and its strategy language, which has been recently incorporated as an official feature. Rewriting strategies and even a primitive version of this language have already been used to represent membrane systems~\cite{membrane,andreiCL06}, but these prototypes have been less evolved than other encodings of membrane systems in rewriting logic~\cite{membraneJournal,exmcPsystem}. The main advantage of our approach is that it transforms a membrane specification to a full-fledged strategy-controlled rewriting system with clearly generated strategy expressions that can be used to faithfully simulate and analyze the evolution of any configuration of the system.
In particular, the authors of the first strategy-based prototype pointed out the difficulty to apply analysis tools to it like the model checker, already used in their first work~\cite{exmcPsystem}. This is solved for free in our case by using the model checkers for strategy-controlled systems and its \emph{opaque strategies} features.
Other advantages are a wider temporal logic support for model checking including LTL, CTL*, $\mu$-calculus, and in general any other logic that would be implemented for strategy-controlled Maude programs; its efficiency shown in~\cref{sec:performance}, since they are now executed by the Maude C++ engine; and its extensibility and adaptability, illustrated in~\cref{sec:variations}.
As mentioned in~\cref{sec:relatedwork}, there are other examples of model-checking tools for membrane systems not based on rewriting, like k\textsc{PWorkbench}~\cite{modelCheckingMembranes} supporting LTL and CTL properties of kernel P systems. In fact, this tool lets the user express priorities and other control mechanisms with ad-hoc strategies.

	As future work, other well-known extensions of membrane systems can be implemented like antiport rules, nonintegral object multiplicities, etc. Tissue-like, neural-like, or probabilistic P systems could also be implemented with broader changes. Moreover, the simulation and verification capabilities can be enhanced with other symbolic techniques supported by Maude like narrowing and SMT solving.

\paragraph{Declaration of competing interest}

	The authors declare that they have no known competing financial interests or personal relationships that could have appeared to influence the work reported in this paper.

\paragraph{Acknowledgements}

	Research partially supported by MCI Spanish projects \emph{TRACES} (TIN2015-67522-C3-3-R) and ProCode-UCM (PID2019-108528RB-C22). Rubén Rubio is partially supported by MU grant FPU17/02319.
	
\appendix

\section{Proofs}

\setcounter{prop}{0}

In the following, we will write $t \to^{\alpha} t'$ to mean that $t'$ is a result of applying the strategy $\alpha$ on the term $t$, $w \to_{A_k} w'$ to say that the multiset $w$ is rewritten to the multiset $w'$ by the multiset of rules $A_k$ (typically in a membrane $M_k$), and $C \to_A C'$ to state that the membrane configuration $C'$ follows from $C$ by an evolution step with a choice $A = (A_k)_{k=1}^n$ for each membrane $M_k$.

\begin{lemma} \label{lemma:el}
	Given a membrane system $\Pi = (O, \mu, w_1, \ldots, w_n, R_1, \ldots, R_n, i_o)$ and its rewriting logic representation $\mathcal R = (\Sigma, E, R)$ with strategies described in~\cref{sec:repr}, there is a bijective mapping $T$ between (1) objects $O$ and ground \texttt{Obj} terms, (2) multisets of objects and ground terms of sort \texttt{ObjSoup}, (3) target messages and ground terms of sort \texttt{TargetMsg}, (4) multisets of targets and ground terms of sort \texttt{TargetSoup}, (5) membrane configurations, multisets of membrane configurations, and heterogeneous multisets (including objects, targets, and membranes) and ground terms of sort \texttt{Membrane}, \texttt{MembraneSoup}. and \texttt{Soup}, respectively, and (5) evolution rules and ground rewrite rules from \texttt{ObjSoup} to \texttt{TargetSoup}. Moreover, the multiset $w$ is rewritten to $w'$ by an evolution rule $r$ iff $T(w)$ is rewritten to $T(w')$ by its corresponding rewrite rule $T(r)$. And $w$ is rewritten to $w'$ by a multiset $A$ of evolution rules iff $T(w)$ is rewritten to $T(w')$ by applying the translation of the rules in $A$ in any order.
	
	Assuming that \texttt{handleMembrane} is terminating, and for $1 \leq k \leq n$ and for any multisets $w$ and $w'$ of objects
\begin{equation}
	w \to_{A_k} w' \; \text{ iff } \; T(w) \to^{\texttt{handleMembrane($M_k$)}} T(w'),  \label{lemma:cond}
\end{equation}
then \texttt{mpr} is terminating and for any configurations $C$ and $C'$, $C \to_A C'$ iff $T(C) \to^\texttt{mpr} T(C')$.
\end{lemma}

\begin{proof}
	The specification described in~\cref{sec:repr} consists of a common infrastructure and some user-defined additions for the particular $\Pi$. These include:
\begin{itemize}
\item A name for each membrane is defined as a constant of sort \texttt{MembraneName}. While the definition of membrane system in~\cref{sec:mems} does not refer to membrane names, membranes are still numbered from $1$ to $n$, so a simple bijection $k \leftrightarrow M_k$ relates both nomenclatures.
	\item Each object $o \in O$ of the membrane system is represented as a term of sort \texttt{Obj}, which also includes the predefined symbol \texttt{delta} for $\delta$. In principle, a constant should be defined for each object, but using terms with more complex structure is not a problem. Hence, $O$ is in bijection with the set $T_{\Sigma,\texttt{Obj}}(\emptyset)$ of all ground terms of sort \texttt{Obj} by construction.
	
		The generic part of the rewriting logic representation includes a sort \texttt{ObjSoup} with \texttt{Obj} as subsort, \texttt{empty} as constant, and the commutative and associative yuxtaposition operator \texttt{\_\_} as data constructor. Ground terms of this sort are exactly multisets of objects, since they are strings of objects identified regardless of their order, with the identity element \texttt{empty} in the role of the empty multiset. Similarly, there is also a bijection between target messages, which are pairs with a multiset of objects and a target annotation, and ground terms of the \texttt{TargetMessage} sort. \texttt{TargetSoup} terms are made out of \texttt{TargetMessage} terms as \texttt{ObjSoup} terms were built from \texttt{Obj} terms, so they are also in bijection with multisets of targets.
		
	\item Each evolution rule is defined as a labeled rewrite rule from a term of sort \texttt{ObjSoup} to a term of sort \texttt{TargetSoup}. Since there is a bijection between the sets of ground terms of these sorts and the multisets of objects and targets, respectively, evolution rules and rewrite rules of that form are also in one-to-one correspondence. Moreover, applying an evolution rule to a multiset of objects is equivalent to applying the corresponding rewrite rule to a \texttt{Soup} term, because the matching objects will be removed from the \texttt{Soup}-multiset and the new targets will be added. When a multiset of evolution rules is applied, the union of their left-hand sides is replaced in the multiset by the union of their right-hand sides. Although their equivalent rewrite rules are applied sequentially, they will produce the same result, since they will replace their disjoint left-hand sides by a \texttt{TargetSoup} term that cannot be modified by other any such rewrite rule. Indeed, even though these messages contain multisets of objects, they have been declared as frozen so that rules cannot be applied inside them. We have also required that \texttt{here} targets should always be used instead of spare objects in the right-hand side to achieve this effect. Since they operate on disjoint parts of the \texttt{Soup} term, the order in which they are applied is irrelevant.
	
	\item Every rewrite rule $r \in R_k$ defined as in the previous item is assigned to its membrane $M_k$ by including it in the definition of the strategy \texttt{membraneRules($M_k$)}.
\end{itemize}
The structure $\mu$ and the initial contents $w_k$ of the membrane system are not specified within the Maude module, but as part of the initial term $t$ on which the strategies are to be applied. This term of sort \texttt{Membrane} is built with the \texttt{<\_|\_>} constructor from a membrane name and a multiset of objects, targets, and nested membranes represented by a \texttt{Soup} term. Reasoning inductively, we can simultaneously prove that \texttt{Membrane} and \texttt{MembraneSoup} terms are in one-to-one correspondence to membrane configurations and multisets of them.
The initial contents $w_k$ of the membrane $k$ can be obtained as the restriction to $O$ of the \texttt{Soup}-multiset within the pair for $M_k$. Moreover, the structure of the membrane $\mu$ (as a tree) can be obtained inductively by looking at the multiset of membranes while adding their labels as children.

	At this point, we have to check that applying the \texttt{mpr} strategy to the \texttt{Membrane} term $T(C)$ yields all possible evolution steps from the membrane configuration $C$. Remember that \texttt{mpr} is defined as \texttt{(visit-mpr ; \skywd{amatch} TM ) ; communication ; (dis !)} and that an evolution step consists of three consecutive phases. We claim that these phases match the three concatenated strategies in the \texttt{mpr} definition, which are executed consecutively (the results of any of them are continued by the next one) by the semantics of the strategy composition operator. Let us first look at the second and third phases:
\begin{enumerate}
	\setcounter{enumi}1
	\item In the second phase, \texttt{out} messages are transferred to the enclosing membrane, \texttt{in $M$} messages are moved to the nested membrane $M$, and \texttt{here} messages are left in the current one. For a single message, this is clearly the meaning of the \texttt{out}, \texttt{in}, and \texttt{here} rules. The \texttt{communication} strategy applies them repeatedly until no more can be applied, so it fulfills the requirements of the second phase. In effect, the strategy is defined as \texttt{(in | out | here) !}, where \texttt{!}\ means the successive application of its argument until it fails, and its argument \texttt{in | out | here} is the nondeterministic application of any of these rules. This disjunction will only fail when none is applicable, and this only happens when no valid target is in the configuration.
	\item In the third phase, membranes containing the $\delta$ symbol are dissolved. The semantics of the \texttt{dis} rule is clearly the dissolution of a non-skin membrane, and the \texttt{!} operator repeats \texttt{dis} until it fails, i.e., until no more $\delta$ symbols are present in a non-skin configuration.  
\end{enumerate}
Finally, we must prove the correctness of the first step \texttt{visit-mpr ; \skywd{amatch} TM}. The test with the variable \texttt{TM} of sort \texttt{TargetMsg} as pattern is a way of checking that at least a rule has been applied in the whole system, as required by the definition. Since the test variant is \skywd{amatch}, it will try to match \texttt{TM} everywhere, so it will succeed iff there is a target message in the configuration, or equivalently, whenever a rule has been applied. Remember that we assume that explicit \texttt{here} targets are used in the encoding of evolution rules instead of spare objects. On the other hand, \texttt{visit-mpr} applies \texttt{handleMembrane} to the multiset of objects in the membrane, and the strategy \texttt{nested-mpr} to the multiset of nested membranes, which it takes as argument. These are the semantics of the \skywd{matchrew} combinator, which matches the pattern on the subject term, and rewrites the subterms matched by some of the pattern variables with some given strategies. The \texttt{nested-mpr} strategy does nothing (\skywd{idle}) if there are no nested membranes in its argument, and otherwise takes one \texttt{M} out of the multiset, applies \texttt{visit-mpr} to it, and continues recursively with the rest \texttt{MS}. In summary, the strategy \texttt{visit-mpr} applies \texttt{handleMembrane} to its object multiset and \texttt{visit-mpr} to every nested membrane, so that \texttt{handleMembrane} is applied to the multiset of objects of every membrane. Since the assumption~\ref{lemma:cond} tells that \texttt{handleMembrane} executes a maximal parallel rewriting step when applied to the multiset of objects of a membrane, we conclude that \texttt{visit-mpr} applies a maximal parallel step on every membrane, as required for the first stage of the evolution step.

	With regard to termination, take into account that the execution of the arguments of the \texttt{!}\ operator in the second and third phases decreases the number of targets or $\delta$ symbols in the configuration, which are finite. Hence, these arguments will eventually fail and the normalization operator \texttt{!}\ will eventually stop. Similarly, the number of membranes in the \texttt{nested-mpr} argument decreases with each call, until zero when no action is taken. For \texttt{visit-mpr}, \texttt{handleMembrane} is applied only once per call, and \texttt{visit-mpr} itself is only called recursively as many times as membranes are in the system, but this number is finite.
\end{proof}

After this lemma, we will usually identify the elements of the membrane system and the rewriting-logic entities that represent them, since they are in one-to-one correspondence.

\begin{prop}
	The strategy \texttt{mpr} executes a maximal parallel evolution step without priorities when \texttt{membraneRules($M$)} is defined as the disjuntion of all rules for $M$, i.e., under these conditions, $C \to_A C'$ iff $T(C) \to^{\texttt{mpr}} T(C')$ for any configurations $C$ and $C'$.
\end{prop}

\begin{proof}
	Thanks to~\cref{lemma:el}, we only have to prove that \texttt{handleMembrane($M_k$)} behaves as expected for every membrane $M_k$. Remember that this strategy is defined in this case as:
\begin{lstlisting}[mathescape]
sd handleMembrane(MN) := inner-mpr(MN) .
sd inner-mpr(MN) := membraneRules(MN) ! .
sd membraneRules($M_k$) := $r_1$ | $\cdots$ | $r_n$ .
\end{lstlisting}
where $r_1, \ldots, r_n$ are all the rules belonging to $M_k$. According to the semantics of the disjunction combinator, an execution of \texttt{membraneRules($M_k$)} is the application of one of the $r_i$ rules nondeterministically chosen. The meaning of \texttt{inner-mpr} is the repeated application of \texttt{membraneRules} until it fails, as follows from the semantics of the normalization operator \texttt{!}.  Only the rules belonging to $M_k$ will be applied since the membrane name is given as an argument to the \texttt{membraneRules} strategy and only the definition for the given name will be executed. The strategy \texttt{inner-mpr} is terminating since the rules $r_i$ remove at least one object in the multiset, they do not introduce any object without a target, and the number of objects is finite. Let $A : R_k \to \N$ be the multiset of evolution rules whose equivalent rewrite rules have been applied by \texttt{inner-mpr}. We claim that $A$ can be applied and is maximal. We already know from the lemma that the result of the parallel application is the result of the strategy, and that the order in which the rules have been applied is immaterial.

	It is clear that $A$ can be applied because the union of the left-hand sides of its rules must have been present in the multiset to trigger the execution of the rewrite rules. Suppose $A$ is not maximal, then we could add an extra rule $r$ to the multiset, or equivalently apply $r$ after all other rules in $A$ have been applied. However, \texttt{membraneRules(MN)} has failed after executing the last rule in $A$, meaning that no rule in the disjunction could have been applied. Hence, by contradiction, $A$ is maximal.
	
\end{proof}

\begin{prop}
	The strategy \texttt{mpr} executes a maximal parallel evolution step with weak priorities when \texttt{membraneRules($M$)} is defined as indicated above this statement in~\cref{sec:repr}.
\end{prop}

\begin{proof}
	Like in the previous proposition, we have to prove that \texttt{handleMembrane} satisfies the requirements of~\cref{lemma:el}. In this case, \texttt{membraneRules} is also repeated until it fails, but now it has a different definition.  Given a generator set of priorities $P_k \subseteq R_k \times R_k$ for the membrane $M_k$, and being $\rho_k$ its transitive closure, the strategy is defined recursively from $R_k$ as indicated in~\cref{sec:repr}. Remember that $A$ is admissible in this case if for every $r \in R_k$ either $A(r') = 0$ for all $r >_{\rho_k} r'$ or $A[r'/0]_{r >_{\rho_k} r'} + \{r\}$ cannot be applied.
	
	First, we claim that $w'$ is a result of \texttt{handleMembrane($M_k$)} on $w$ iff there is a rule $r \in R_k$ such that no rule $r' >_{\rho_k} r$ can be applied and $w'$ is the result of applying $r$ to $w$. According to the procedure for constructing the weak-priority strategy, every occurrence of a rule $r$ is preceded by its immediate predecessors in the order as $(r_1 | \cdots | r_n) \;\skywd{or-else}\; r$. Hence, by the semantics of the operator $\alpha \;\skywd{or-else}\; \beta \equiv \alpha \;\texttt?\; \skywd{idle} \;\texttt:\; \beta$, $r$ cannot be applied if any one of these $r_1$ to $r_n$ are applicable. Inductively, these $r_i$ are also guarded by their predecessors, so if any one of them can be applied, $r_i$ cannot be executed, but neither can $r$. Indeed, the solution of the ancestor is a solution of the \skywd{or-else} ending in $r_i$, and this is a solution of the left-hand side of the \skywd{or-else} whose right-hand side is $r$, so that $r$ is not executed. Every rule in $R_k$ is included in \texttt{handleMembrane($M_k$)}, because every one is reachable from the minimal elements of the order, and so it should have been inserted by the first or second points of the procedure. If it is a maximal element it will be executed immediately;  otherwise, after the rules with greater priority have been discarded. Moreover, the strategy applies a single rule, because only one of the branches in the multiple disjunction combinators is chosen on each occasion, and once the left-hand side of an \skywd{or-else} succeeds, all enclosing \skywd{or-else} will finish with the \skywd{idle} of the positive branch of the equivalent conditional expression without executing any other rule.
		
	Using the arguments already given for the proof of the case without priorities in the previous proposition, we know that \texttt{handleMembrane} is terminating, $A$ can be applied, and the results of the parallel application of $A$ and \texttt{handleMembrane($M_k$)} coincide, regardless of the order in which rules have been applied. We should then prove that $A$ is admissible and maximal. Suppose it is not admissible, so for some $r \in R_k$ there is $r'$ such that $r >_{\rho_k} r'$ and $A(r') > 0$, and the choice $A[r''/0]_{r >_{\rho_k} r''} + \{r\}$ can be applied. Since the rules in the multiset can be applied in any order, we could have executed $r'$ after $A[r''/0]_{r >_{\rho_k} r''}$, but at this moment $r$ can be applied too. This contradicts the claim in the previous paragraph, which says that $r'$ cannot be applied if $r >_{\rho_k} r'$ can be applied, so $A$ is admissible. In order to prove the maximality, suppose that $A + \{ r \}$ can be applied, then the rule $r$ should be applicable just after the rules of $A$. However, \texttt{handleMembrane} has failed, and this means that either $r$ cannot be applied or there is an $r'$ with $r' >_{\rho_k} r$ that can be applied. In the first case we have arrived to a contradiction. In the second, we can repeat the argument until we arrive to the maximal element, where the contradiction is unavoidable. Hence, $A$ is maximal.
\end{proof}

\begin{prop}
	The strategy \texttt{mpr} executes a maximal parallel evolution step with strong priorities when \texttt{handleMembrane} is instantiated to the \texttt{strong-mpr} strategy described above this statement in~\cref{sec:repr}.
\end{prop}

\begin{proof}
	According to~\cref{lemma:el}, it is enough to prove that \texttt{strong-mpr} is terminating and satisfies \ref{lemma:cond}.
First, the strategy \texttt{strong-mpr} is terminating as follows from taking the number of objects in the multiset where it is applied as a rank function. Every recursive call is preceded by a rule application, and the former is only executed if the latter succeeds. Rules always reduce the number of objects in the multiset, so the rank function always decreases between recursive calls. This is clear in the basic strategies $\alpha_r$ where the application of $r$ is concatenated with the recursive call. According to the semantics of this combinator, the first one must have succeeded for the second one to be applied. The transformation does not introduce any additional recursive calls, except through basic strategies, so this property is preserved.
	
	Let $A$ be the set of rules applied by \texttt{strong-mpr}, which coincides with the argument of its deepest recursive call. Remember that $A$ is admissible in the strong sense if it is admissible in the weak sense and $A(r) > 0$ implies $A(r') = 0$ for all $r >_{\rho_k} r'$.
The choice clearly satisfies the second condition for any pair of rules $r >_{\rho_k} r'$, because $r$ is added to the argument \texttt{AP} of \texttt{strong-mpr} just after $r$ is applied, it is never removed in a recursive call, and so the \skywd{match} test that guards the application of $r'$ will fail and impede its execution. The recursive call that applies $r$ as well as the previous calls, which may not have $r$ in its argument \texttt{AP}, will not execute any rules after the recursive call for $r$ has returned, because the success of $r$ will finish the enclosing \skywd{or-else} and strategy calls in cascade, as mentioned in the previous proposition. The execution of \texttt{strong-mpr} always succeeds because of the \skywd{idle} added in the fourth step of the procedure.
The proof that the choice obtained with the strategy is admissible in the weak sense and maximal is the same as in the previous case, taking into account that their strategies share the same structure, even though the next iteration is repeated by \texttt{inner-mpr} in the first case and by a recursive call in this case.
\end{proof}

\bibliographystyle{elsarticle-harv}

\end{document}